\documentclass[preprint,5p,number,sort&compress,times,a4paper]{elsarticle}

\usepackage[T1]{fontenc}
\usepackage[fleqn]{amsmath}
\usepackage{amssymb,amsfonts}
\usepackage{booktabs} 
\usepackage{mathtools}
\usepackage{microtype}
\usepackage{algorithmicx}
\usepackage{algcompatible}
\usepackage{pgfplots}
\pgfplotsset{compat=newest}
\usepackage{tikz}
\usetikzlibrary{shapes,arrows,decorations,backgrounds,positioning,calc,patterns,decorations.pathmorphing,decorations.markings,spy,backgrounds}
\usepackage{import}
\newlength\figureheight
\newlength\figurewidth
\usepackage{nicefrac}

\usepackage[english]{babel}
\usepackage{blindtext}

\newtheorem{thm}{Theorem}
\newtheorem{prop}{Proposition}
\newtheorem{lem}{Lemma}
\newdefinition{rem}{Remark}
\newdefinition{assum}{Assumption}
\newdefinition{alg}{Algorithm}
\newdefinition{crit}{Criterion}
\newdefinition{defn}{Definition}
\newdefinition{cor}{Corollary}
\newproof{pf}{Proof}

\newlength{\hatchspread}
\newlength{\hatchthickness}
\newlength{\hatchshift}
\newcommand{\hatchcolor}{}
\tikzset{hatchspread/.code={\setlength{\hatchspread}{7pt}},
	hatchthickness/.code={\setlength{\hatchthickness}{#1}},
	hatchshift/.code={\setlength{\hatchshift}{#1}},% must be >= 0
	hatchcolor/.code={\renewcommand{\hatchcolor}{#1}}}
\tikzset{hatchspread=3pt,
	hatchthickness=0.4pt,
	hatchshift=0pt,% must be >= 0
	hatchcolor=black}
\pgfdeclarepatternformonly[\hatchspread,\hatchthickness,\hatchshift,\hatchcolor]% variables
{nwl}% name
{\pgfqpoint{\dimexpr-2\hatchthickness}{\dimexpr-2\hatchthickness}}% lower left corner
{\pgfqpoint{\dimexpr\hatchspread+2\hatchthickness}{\dimexpr\hatchspread+2\hatchthickness}}% upper right corner
{\pgfqpoint{\dimexpr\hatchspread}{\dimexpr\hatchspread}}% tile size
{% shape description
	\pgfsetlinewidth{\hatchthickness}
	\pgfpathmoveto{\pgfqpoint{0pt}{\dimexpr\hatchspread+\hatchshift}}
	\pgfpathlineto{\pgfqpoint{\dimexpr\hatchspread+0.15pt+\hatchshift}{-0.15pt}}
	\ifdim \hatchshift > 0pt
	\pgfpathmoveto{\pgfqpoint{0pt}{\hatchshift}}
	\pgfpathlineto{\pgfqpoint{\dimexpr0.15pt+\hatchshift}{-0.15pt}}
	\fi
	\pgfsetstrokecolor{\hatchcolor}
	\pgfusepath{stroke}
}
\pgfdeclarepatternformonly[\hatchspread,\hatchthickness,\hatchshift,\hatchcolor]% variables
{nwld}% name
{\pgfqpoint{\dimexpr-2\hatchthickness}{\dimexpr-2\hatchthickness}}% lower left corner
{\pgfqpoint{\dimexpr\hatchspread+2\hatchthickness}{\dimexpr\hatchspread+2\hatchthickness}}% upper right corner
{\pgfqpoint{\dimexpr\hatchspread}{\dimexpr\hatchspread}}% tile size
{% shape description
	\pgfsetlinewidth{\hatchthickness}
	\pgfpathmoveto{\pgfqpoint{0pt}{\dimexpr\hatchspread+\hatchshift}}
	\pgfpathlineto{\pgfqpoint{\dimexpr\hatchspread+0.15pt+\hatchshift}{-0.15pt}}
	\ifdim \hatchshift > 0pt
	\pgfpathmoveto{\pgfqpoint{0pt}{\hatchshift}}
	\pgfpathlineto{\pgfqpoint{\dimexpr0.15pt+\hatchshift}{-0.15pt}}
	\fi
	\pgfsetstrokecolor{\hatchcolor}
	\pgfsetdash{{1pt}{1pt}}{0pt}% dashing cannot work correctly in all situation this way
	\pgfusepath{stroke}
}
\pgfdeclarepatternformonly[\hatchspread,\hatchthickness,\hatchshift,\hatchcolor]% variables
{nel}% name
{\pgfqpoint{\dimexpr-2\hatchthickness}{\dimexpr-2\hatchthickness}}% lower left corner
{\pgfqpoint{\dimexpr\hatchspread+2\hatchthickness}{\dimexpr\hatchspread+2\hatchthickness}}% upper right corner
{\pgfqpoint{\dimexpr\hatchspread}{\dimexpr\hatchspread}}% tile size
{% shape description
	\pgfsetlinewidth{\hatchthickness}
	\pgfpathmoveto{\pgfqpoint{\dimexpr\hatchshift-0.15pt}{-0.15pt}}
	\pgfpathlineto{\pgfqpoint{\dimexpr\hatchspread+0.15pt}{\dimexpr\hatchspread-\hatchshift+0.15pt}}
	\ifdim \hatchshift > 0pt
	\pgfpathmoveto{\pgfqpoint{-0.15pt}{\dimexpr\hatchspread-\hatchshift-0.15pt}}
	\pgfpathlineto{\pgfqpoint{\dimexpr\hatchshift+0.15pt}{\dimexpr\hatchspread+0.15pt}}
	\fi
	\pgfsetstrokecolor{\hatchcolor}
	\pgfusepath{stroke}
}
\pgfdeclarepatternformonly[\hatchspread,\hatchthickness,\hatchshift,\hatchcolor]% variables
{neld}% name
{\pgfqpoint{\dimexpr-2\hatchthickness}{\dimexpr-2\hatchthickness}}% lower left corner
{\pgfqpoint{\dimexpr\hatchspread+2\hatchthickness}{\dimexpr\hatchspread+2\hatchthickness}}% upper right corner
{\pgfqpoint{\dimexpr\hatchspread}{\dimexpr\hatchspread}}% tile size
{% shape description
	\pgfsetlinewidth{\hatchthickness}
	\pgfpathmoveto{\pgfqpoint{\dimexpr\hatchshift-0.15pt}{-0.15pt}}
	\pgfpathlineto{\pgfqpoint{\dimexpr\hatchspread+0.15pt}{\dimexpr\hatchspread-\hatchshift+0.15pt}}
	\ifdim \hatchshift > 0pt
	\pgfpathmoveto{\pgfqpoint{-0.15pt}{\dimexpr\hatchspread-\hatchshift-0.15pt}}
	\pgfpathlineto{\pgfqpoint{\dimexpr\hatchshift+0.15pt}{\dimexpr\hatchspread+0.15pt}}
	\fi
	\pgfsetstrokecolor{\hatchcolor}
	\pgfsetdash{{1pt}{1pt}}{0pt}% dashing cannot work correctly in all situation this way
	\pgfusepath{stroke}
}

\begin{document}
\begin{frontmatter}
\title{Stabilizing predictive control with persistence of excitation for constrained linear systems}

\author[Sheffield]{Bernardo A. Hernandez\corref{cor1}\fnref{fn1}}
\ead{bahernandezvicente1@sheffield.ac.uk}
\author[Sheffield]{Paul A. Trodden}
\ead{p.trodden@sheffield.ac.uk}

\address[Sheffield]{Department of Automatic Control \& Systems
Engineering, University of Sheffield, Mappin Street, Sheffield S1
3JD, UK} 

\cortext[cor1]{Corresponding author. Tel: +44(0)114 2225250. Fax: +44(0)114 2225683.}
\fntext[fn1]{B. A. Hernandez is
supported by the Chilean Ministry of Education
(CONICYT PFCHA/Concurso para Beca de Doctorado en el Extranjero - 72150125).}


\begin{abstract} % Abstract of not more than 200 words.
A new adaptive predictive controller for constrained linear systems is presented. The main feature of the proposed controller is the partition of the input in two components. The first part is used to persistently excite the system, in order to guarantee accurate and convergent parameter estimates in a deterministic framework. An MPC-inspired receding horizon optimization problem is developed to achieve the required excitation in a manner that is optimal for the plant. The remaining control action is employed by a conventional tube MPC controller to regulate the plant in the presence of parametric uncertainty and the excitation generated for estimation purposes. Constraint satisfaction, robust exponential stability, and convergence of the estimates are guaranteed under design conditions mildly more demanding than that of standard MPC implementations.
\end{abstract}
\end{frontmatter}

\section{Introduction}
Model predictive control (MPC) is an advanced control technique that handles constraints explicitly and optimizes system performance online \cite{Rawlings2014}. However, because the synthesis of an MPC controller requires a model of the plant being controlled, any guarantees that conventional MPC provides are \emph{nominal}; in practice, stability and performance rely on the model being a sufficiently accurate representation of the plant. Robust forms of MPC (such as \cite{Kothare1996,Chisci2001,Mayne2005}) seek, therefore, to establish guarantees when the uncertainty in the system (including modelling error) can be bounded. Yet robust MPC takes a worst-case approach to the problem, and predictions are usually made using some fixed nominal model \cite{Mayne2005}, thus closed-loop performance can be poor.

Adaptive MPC (AMPC) aims to overcome some of these drawbacks by identifying (and providing the MPC controller with) a more accurate model of the system during operation. Albeit several approaches have been proposed, e.g. \cite{Fukushima2007,Adetola2009,Wang2014a,Aswani2013,Tanaskovic2014,Fan2016,Genceli1996,Vuthandam1997a,Shouche1998,Marafioti2013}, AMPC remains to a large extent an open problem \cite[Section~3.1]{Mayne2014}. One of the reasons for this is the duality \cite{Goodwin1984} of the optimal control problem, in which the objectives of achieving sufficient excitation of the system for a successful (closed-loop) identification and satisfactory regulation are competing.

Different AMPC approaches place different emphasis on these competing objectives. In \cite{Fukushima2007,Adetola2009,Wang2014a,Aswani2013,Tanaskovic2014,Fan2016}, the main concern is the control objective, and different robust approaches (such as min-max optimization \cite{Adetola2009} and constraint tightening \cite{Aswani2013,Tanaskovic2014,Fan2016}) are employed in order to ensure robust constraint satisfaction and stability. The main assumption common to these approaches is that a bound on the initial modelling error is known, and that it does not increase over time; the latter is achieved by allowing parameter estimates to be updated only when the closed-loop output data is informative enough \cite{Ljung1999}, but there are no guarantees that this will occur. Therefore, while attaining desirable system theoretic properties, these approaches assume, rather than guarantee, sufficient excitation of the unknown system for accurate identification. On the other hand, a group of papers that focus on guaranteeing sufficient excitation, albeit at the expense of constraint satisfaction and stability guarantees, is \cite{Genceli1996,Vuthandam1997a,Shouche1998,Marafioti2013}. In these approaches, a nominal (non-robust) MPC optimization problem is augmented with a constraint that forces the input sequence to be persistently exciting (PE) \cite{Ljung1999,Goodwin1984}. In \cite{Marafioti2013}, the receding-horizon principle of MPC is explicitly taken into account in the design of the PE input, which allows for a recursive feasibility guarantee with respect to the PE constraint; however, constraint satisfaction and stability of the closed-loop system are merely assumed.

Some approaches have been proposed to address both objectives simultaneously, and achieve control guarantees while ensuring sufficient excitation. In \cite{Ferramosca2013a}, the system states are driven to a region of the state space wherein identification experiments can be performed safely (i.e. without constraint violation). The approach is, however, limited to open-loop stable linear time invariant (LTI) systems and, moreover, system uncertainty is entirely neglected during the transient; constraint satisfaction and convergence to the target region are not guaranteed during this phase. In \cite{Weiss2014}, robust set invariance concepts are employed in order to guarantee constraint satisfaction by the trajectories of the uncertain controlled system. Excitation is promoted via an augmented cost function in the MPC problem; this results in a non-convex optimization problem, albeit the system model and constraints are linear and the regulation objective is quadratic.

In this paper we propose a new and simple solution to the AMPC problem for linear constrained systems, that achieves guarantees of stability, constraint satisfaction and persistence of excitation. Our approach is to decouple the objectives of regulation and excitation, thereby making their fulfilment more straightforward; we show that this can be achieved by partitioning the control input into a regulatory part and an exciting part, and then deploying conventional tube-based MPC \cite{Mayne2005} in order to control the uncertain excited system. Robust stability and constraint satisfaction are guaranteed even if the plant is open-loop unstable (c.f. \cite{Ferramosca2013a,Tanaskovic2014}) and linear time varying (LTV). Convergence of the parameter estimates is achieved by inclusion of PE-type constraints, similar to \cite{Genceli1996,Marafioti2013}, in a separate optimization problem built specifically for the exciting input design; however, the main MPC problem remains convex, unlike in \cite{Weiss2014}.

The main drawback of the proposed approach is its conservativeness and the strength of our assumptions: in particular, the control guarantees rely on some a-priori knowledge of the uncertain plant (its order and bounds on uncertainty). On the other hand, the design complexity of our approach is lower than in \cite{Adetola2009,Fukushima2007,Wang2014a,Tanaskovic2014}, and comparable to \cite{Aswani2013,Ferramosca2013a,Weiss2014}; moreover, we provide some insights into how the assumptions might be met in practice.

A preliminary version of this approach appeared in \cite{Hernandez2016}. Several modifications and additional contributions have been included in this paper, amongst them (i) the required excitation is guaranteed to be transmitted from the exciting part to the whole input; (ii) sufficient conditions for the existence of a stable linear feedback gain for the true plant are given instead of assumed; and (iii) the problem of prediction model update is tackled by a set of a-posteriori (online) redesign approaches, instead of an overly robustified initial design. The latter simplifies the design procedure and relaxes many of the assumptions, thus making the proposed controller applicable for larger model uncertainties.

The paper is organized as follows. In Section \ref{sec.prem} the preliminaries of the problem are presented. Section \ref{sec.rc} briefly describes the robust MPC approach and its related assumptions, while Section \ref{sec.pe} develops a novel MPC-like constrained optimization for the purpose of excitation. Section \ref{sec.pmu} discusses three possible approaches for allowing an update of the prediction model, and a numerical example is shown in Section \ref{sec.numerical} to illustrate the performance of the proposed AMPC controller. 

\section{Preliminaries} \label{sec.prem}
\subsection{Notation}
For $\mathbb{C},\mathbb{D}\subset\mathbb{R}^{n}$, $\mathbb{C}\oplus \mathbb{D}$ and $\mathbb{C}\ominus \mathbb{D}$ are, the Minkowski sum and Pontryagin difference respectively \cite{Kolmanovsky}.
A compact set that contains the origin is a $\mathcal{C}$-set and a $\mathcal{PC}$-set if it contains the origin in its interior.
The null matrix is $\boldsymbol{0}$, and the identity is $\mathcal{I}$ (the dimension will be clear from context).
For $x\in\mathbb{R}^{n}$ and $Q\in\mathbb{R}^{n\times n}$, $||x||^2_Q$ is shorthand for $x^{\top} Q x$.
For a time signal $\phi(\cdot)$, the sequence of its values up to time instant $i$ is $\left\{\phi(i)\right\}=\left\{\phi(0),\phi(1),\cdots,\phi(i)\right\}$.

\subsection{Model dynamics and constraints}
Consider the problem of regulating the uncertain LTV system
\begin{equation} \label{eq.1.1}
x(i+1)=A(i)x(i)+B(i)u(i),
\end{equation}
where $x(i)\in\mathbb{R}^{n}$ and $u(i)\in\mathbb{R}^{m}$ are respectively the state and input vectors at the current time instant, and $x(i+1)\in\mathbb{R}^{n}$ is the state vector at the subsequent time. The state and input matrices are uncertain but assumed to reside, at all times, in a compact set, i.e. $\left[A(i)\:\:B(i)\right]\in\mathcal{M}\subset\mathbb{R}^{(n)\times(n+m)}$. With a slight abuse of notation, the time dependency of the state and input matrices is neglected in the rest of the paper. Additionally, the states and inputs are subject to the following constraints
\begin{equation} \label{eq.1.4}
x(i)\in\mathbb{X}\subset\mathbb{R}^{n},\quad u(i)\in\mathbb{U}\subset\mathbb{R}^{m},\quad\forall i\geq 0.
\end{equation}
It is common that an initial guess of the plant parameters, say $\bigl(\bar{A},\bar{B}\bigr)$, is available, thus we can recast \eqref{eq.1.1} in nominal form
\begin{equation}\label{eq.1.2}
x(i+1)=\bar{A}x(i)+\bar{B}u(i)+w_{p}(i),
\end{equation}
where $w_{p}(i)$ is state/input dependent uncertainty arising from the model mismatch (i.e., a parametric uncertainty). The robust dual controller proposed in this paper requires the following assumptions to hold.
\begin{assum} \label{assum.sets}
$\mathbb{X}$ and $\mathbb{U}$ are convex $\mathcal{PC}$-sets.
\end{assum}
\begin{assum}\label{assum.punc}
There exists a $\mathcal{C}$-set $\mathbb{W}_{p}$ such that $w_{p}\in\mathbb{W}_{p}$ for all $\left(x,u,\left[A\:\:B\right]\right)\in\mathbb{X}\times\mathbb{U}\times\mathcal{M}$ and nominal model $\bigl(\bar{A},\bar{B}\bigr)$.
\end{assum}
\begin{assum}\label{assum.stab}
The pair $\left(A(i),B(i)\right)$ is stabilizable for all $i\geq0$. Furthermore, $\bar{A}_{K}=\bar{A}+\bar{B}K$ is Schur for some $K\in\mathbb{R}^{m\times n}$.
\end{assum}

\begin{rem}
Assumptions~\ref{assum.sets} and \ref{assum.punc} are tightly related. If $\mathbb{X}$ or $\mathbb{U}$ are unbounded (e.g., loose state constraints), Assumption~\ref{assum.punc} cannot be met.
\end{rem}
\begin{rem}
The set $\mathbb{W}_{p}$ is a conservative bound for a state/input dependent uncertainty, and the main source of conservatism of the proposed approach. However, a similar set is implicitly characterized by the min-max approaches in \cite{Adetola2009,Tanaskovic2014} and explicitly defined in \cite{Aswani2013} by the set that contains the values of the unmodeled dynamics. Furthermore, given the generality of Assumption~\ref{assum.punc}, the set $\mathbb{W}_{p}$ can be easily computed with minimal knowledge of $\mathcal{M}$. In addition, $\mathcal{M}$ does not need to be convex.
\end{rem}

\subsection{System identification} \label{sec.si}
The proposed setting diverges from the classical framework in system identification problems: full state measurement is assumed and no noise is considered (hence all variables are deterministic). Despite this, a standard approach is employed for the purpose of estimating the true model from measured data $\left(x(i),u(i)\right)$. The argument behind this is that all the guarantees associated to the proposed dual controller depend on its robustness features, which can be maintained under the inclusion of measurement (or process) noise and state estimation errors (similar to the output-feedback MPC approach in \cite{Mayne2006}).

\subsubsection{Parameter estimation algorithm}
Consider the following predictor for the plant in \eqref{eq.1.1}:
\begin{subequations} \label{eq.1.6}
\begin{alignat}{2}
\hat{x}^{\top}(i)&=\phi^{\top}(i)\hat{\theta}(i-1)&&\in\mathbb{R}^{n} \label{eq.1.6.1}\\
\phi^{\top}(i)&=\left[x^{\top}(i-1)\:\:u^{\top}(i-1)\right]&&\in\mathbb{R}^{n+m} \label{eq.1.6.2}\\
\hat{\theta}(i)&=\bigl[\mathcal{A}(i)\:\:\mathcal{B}(i)\bigr]^{\top}&&\in\mathbb{R}^{(n+m)\times(n)}, \label{eq.1.6.3}
\end{alignat}
\end{subequations}
where $\left(\mathcal{A}(i),\mathcal{B}(i)\right)$ are the current estimates of $\left(A,B\right)$. At time $i$, $\hat{\theta}(i)$ is computed following a standard RLS algorithm \cite{Ljung1999}:
\begin{subequations} \label{eq.1.7}
\begin{alignat}{1}
\Delta\hat{\theta}(i)&=E(i)^{-1}\phi(i)\left[x^{\top}(i)-\phi^{\top}(i)\hat{\theta}(i-1)\right]\\
E(i)&=\lambda E(i-1)+\phi(i)\phi^{\top}(i),
\end{alignat}
\end{subequations}
where $\Delta\hat{\theta}(i)=\hat{\theta}(i)-\hat{\theta}(i-1)$ and $\lambda$ is a forgetting factor.

\subsubsection{Convergence of the estimates}
It can be shown that, under mild assumptions, convergence of $\hat{\theta}(i)$ to $\left(A,B\right)$ is guaranteed if the regressor $\phi(i)$ is a strongly persistently exciting (SPE) sequence of order 1 \cite{Goodwin1984}.
\begin{defn}[SPE sequence]\label{defn.spe}
A sequence $\left\{\phi(i)\right\}$, is said to be SPE of order $h\geq1$ at time $i$, if there exists a positive integer $l$ and real numbers $\rho_{0},\rho_{1}>0$ such that,
\begin{subequations}
\begin{alignat}{1}
\rho_{1}\mathcal{I}&>\sum_{j=0}^{l-1}\left(\boldsymbol{\phi}_{i-j}\boldsymbol{\phi}^{\top}_{i-j}\right)>\rho_{0}\mathcal{I} \label{eq.1.8.1}\\
\boldsymbol{\phi}_{i-j}&=\left[\phi(i-j)\:\phi(i-j-1)\:\cdots\:\:\:\phi(i-j-h+1)\vphantom{\phi(i-j)}\right]. \label{eq.1.8.2}
\end{alignat}
\end{subequations}
\end{defn}
Definition~\ref{defn.spe} is equivalent to standard PE definitions \cite{Goodwin1984,Green1985,Ljung1999}, but with the observed time window positioned so that the current time instant $i$ lies at the right-hand end of it. The purpose of this is to facilitate the inclusion of a constraint such as \eqref{eq.1.8.1} in the receding horizon context of MPC.

At time $i$, the regressor vector \eqref{eq.1.6.2} contains not only the past input, but also the past state. Since at time $i-1$ the MPC controller has no influence over the state $x(i-1)$, the direct inclusion of a constraint such as \eqref{eq.1.8.1} might pose feasibility problems. A result from \cite{Green1985} that employs the concept of reachability of linear systems can be used to overcome this.
\begin{defn}[Reachability]\label{defin.sreach}
System \eqref{eq.1.1} is state reachable if the matrix $\mathcal{O}_{s}=\left[B\:\:AB\:\:\cdots\:\:A^{n-1}B\right]$ has full row rank.
\end{defn}
\begin{lem}[PE of output reachable systems \cite{Green1985}] \label{lem.pe1}
The output of any output reachable LTI system of McMillan degree $\nu$ is SPE of order 1, independent of initial conditions, iff the input to the system is SPE of order $\nu+1$.
\end{lem}
\begin{cor}[PE of the regressor]\label{cor.pe1}
Assume that the true plant \eqref{eq.1.1} is state reachable. The sequence $\left\{\phi(i)\right\}$, with $\phi(i)$ defined as in \eqref{eq.1.6.2}, is SPE of order $1$ at time $i$ if the input sequence $\left\{u(i)\right\}$ is SPE of order $n+m$ at time $i$.
\end{cor}

\begin{rem}
Corollary~\ref{cor.pe1} is easily extended to account for time-varying systems via a proper selection of $\rho_{0}$ (see Theorems 4.1 and 4.2 of \cite{Green1985b}).
\end{rem}

\subsection{Dual control problem} \label{sec.prob}
The control problem can be summarized as: regulate the plant described by \eqref{eq.1.1}, respecting the constraints \eqref{eq.1.4}, but using only the available model \eqref{eq.1.2}, while simultaneously producing data that is informative enough for the recursion in \eqref{eq.1.7} to provide converging estimates. Moreover, if possible, use the current converged estimates to updated the nominal model \eqref{eq.1.2} in order to obtain more accurate predictions within the MPC context and hence improve performance of the controlled system. In the following, a dual controller which provides a solution to this problem, and its associated set of assumptions, is described.

\section{Robust control} \label{sec.rc}
\subsection{Input partition}
The central feature of the proposed dual controller is the partition of the input into a regulatory part $\hat{u}$, and an exciting part $\hat{w}$, such that at all time instances the input fed to the plant fulfils $u(i)=\hat{u}(i)+\hat{w}(i)$. The nominal model takes the form,
\begin{equation} \label{eq.1.5}
x(i+1)=\bar{A}x(i)+\bar{B}\hat{u}(i)+w(i),
\end{equation}
where the parametric uncertainty and the exciting part of the input have been lumped into a single disturbance term $w(i)=\hat{B}\hat{w}(i)+w_{p}(i)$. Consider the following constraint partition
\begin{subequations} \label{eq.1.3}
\begin{alignat}{2}
\hat{u}(i)&\in\hat{\mathbb{U}}=\alpha\mathbb{U},&&\quad\forall i\geq 0 \label{eq.1.3.1}\\
\hat{w}(i)&\in\hat{\mathbb{W}}=\mathbb{U}\ominus\hat{\mathbb{U}}=(1-\alpha)\mathbb{U},&&\quad\forall i\geq 0\label{eq.1.3.2}
\end{alignat}
\end{subequations}
with $\alpha\in\left(0,1\right)$. It is clear that satisfaction of \eqref{eq.1.3} guarantees satisfaction of the true input constraint \eqref{eq.1.4}. Moreover, it follows that $\hat{\mathbb{W}}$ is a convex $\mathcal{PC}$-set, hence the set that contains the lumped disturbance, $\mathbb{W}=\bar{B}\hat{\mathbb{W}}\oplus\mathbb{W}_{p}$, is at least a convex $\mathcal{C}$-set \cite{Kolmanovsky}.
\begin{rem}\label{rem.partition}
This architecture has two main purposes: (a) to simplify the control problem to that of the robust regulation of a linear system $\bigl(\bar{A},\bar{B}\bigr)$ in the presence of a bounded additive disturbance $w(i)\in\mathbb{W}$, and (b) to allow the selection of $\hat{u}(i)$ to be independent from that of $\hat{w}(i)$, as long as $\hat{w}(i)\in\hat{\mathbb{W}}$ is guaranteed.
\end{rem}

\subsection{Regulation via tube MPC} \label{sec.tube}
In view of Remark~\ref{rem.partition}, the plant is controlled robustly using a simplified version of conventional tube MPC that is developed in detail in \cite[Chapter~3]{Rawlings2014}. For completeness of exposition, we now recall some standard definitions and present a brief description of the optimal control problem devised in \cite[Chapter~3]{Rawlings2014}.
\begin{defn}[Positive invariant (PI) set]
A set $\mathbb{T}\subset\mathbb{R}^{n}$ is a PI set for the dynamics $x(i+1)=\bar{A}_{K}x(i)$ if $\bar{A}_{K}\mathbb{T}\subseteq\mathbb{T}$.
\end{defn}
\begin{defn}[Robust PI (RPI) set]
A set $\mathbb{S}\subset\mathbb{R}^{n}$ is an RPI set for the dynamics $x(i+1)=\bar{A}_{K}x(i)+w(i)$ with $w(i)\in\mathbb{W}$ if $\bar{A}_{K}\mathbb{S}\oplus\mathbb{W}\subseteq\mathbb{S}$.
\end{defn}
\begin{rem}
A PI set $\mathbb{T}$ is called admissible (for constraints \eqref{eq.1.4} and \eqref{eq.1.3.1}) if $\mathbb{T}\subset\mathbb{X}$ and $K\mathbb{T}\subset\hat{\mathbb{U}}$. The same holds for an RPI set.
\end{rem}
Consider an undisturbed representation of \eqref{eq.1.2}
\begin{equation} \label{eq.1.8}
z(i+1)=\bar{A}z(i)+\bar{B}v(i).
\end{equation}
The control law employed to regulate the disturbed plant is
\begin{equation}\label{eq.1.11}
\hat{u}(i)=\hat{\kappa}(x(i),z(i))=\kappa\left(z(i)\right)+K_{t}\left(x(i)-z(i)\right),
\end{equation} 
where $K_{t}$ is stabilizing for $\bigl(\bar{A},\bar{B}\bigr)$ and $\kappa\left(z(i)\right)$ is the receding horizon control law that stems from a nominal MPC controller designed to stabilize the undisturbed model in \eqref{eq.1.8}, subject to tightened versions of constraints \eqref{eq.1.4} and \eqref{eq.1.3}. The optimal control problem to be solved at each time instant is defined as:
\begin{equation} \label{eq.1.9}
\mathbb{P}_{N}\left(z=z(i)\right):\min_{\boldsymbol{v}}V_{N}\left(z,\boldsymbol{v}\right)
\end{equation}
subject to (for $k=0,\ldots,N-1$):
\begin{subequations}  \label{eq.1.10}
\begin{alignat}{1}
z_{0}&=z \label{eq.1.10.1}\\
z_{k+1}&=\bar{A}z_{k}+\bar{B}v_{k} \label{eq.1.10.2}\\
z_{k}&\in\mathbb{Z}=\mathbb{X}\ominus\mathbb{S} \label{eq.1.10.3}\\
v_{k}&\in\mathbb{V}=\hat{\mathbb{U}}\ominus K_{t}\mathbb{S} \label{eq.1.10.4}\\
z_{N}&\in\mathbb{Z}_{f}\subseteq\mathbb{X}\ominus\mathbb{S} \label{eq.1.10.5}.
\end{alignat}
\end{subequations}
A sub-index is employed in \eqref{eq.1.10} to differentiate predictions from true values, $N$ is the controller prediction horizon and the optimization variable $\boldsymbol{v}$ represents the sequence of control actions throughout the prediction horizon, i.e. $\boldsymbol{v}=\left\{v_{0},v_{1},\ldots,v_{N-1}\right\}$. The cost function $V_{N}\left(z,\boldsymbol{v}\right)$ is defined as the standard finite horizon LQR cost with terminal penalty
\begin{equation} \label{eq.1.13}
\begin{aligned}
V_{N}\left(z,\boldsymbol{v}\right)&=\sum_{k=0}^{N-1}\ell\left(z_{k},v_{k}\right)+V_{f}\left(z_{N}\right)\\
&=\sum_{k=0}^{N-1}\left(||z_{k}||^{2}_{Q}+||v_{k}||^{2}_{R}\right)+||z_{N}||^{2}_{P}.
\end{aligned}
\end{equation}
The set of all the states for which the optimization problem \eqref{eq.1.9}--\eqref{eq.1.10} is feasible is defined as $\mathcal{Z}_{N}$. The solution to \eqref{eq.1.9}--\eqref{eq.1.10} is a sequence of optimal inputs and associated predicted states,
\begin{subequations} \label{eq.1.12}
\begin{alignat}{1}
\boldsymbol{v}^{*}(z(i))&=\left\{v^{*}_{0},v^{*}_{1},\ldots,v^{*}_{N-1}\right\} \label{eq.1.12.1}\\
\boldsymbol{z}^{*}(z(i),\boldsymbol{v}^{*})&=\left\{z_{0},z^{*}_{1},\ldots,z^{*}_{N-1},z^{*}_{N}\right\},
\end{alignat}
\end{subequations}
and the implicit nominal control law is defined as the first control action of the optimal sequence $\kappa\left(z(i)\right)=v^{*}_{0}$.

\subsection{Conventional tube MPC properties}
The following result summarizes the main properties of the robust controller described above (see \cite[Chapter 3]{Rawlings2014} for a detailed proof of Theorem~\ref{thm.rstab}).
\begin{assum}\label{assum.pi}
The set $\mathbb{Z}_{f}$ is an admissible PI set for the dynamics in \eqref{eq.1.8} when in closed-loop with a stabilizing linear gain $K$ (possibly different from $K_{t}$).
\end{assum}
\begin{assum}\label{assum.rpi}
The set $\mathbb{S}\subset{\mathbb{R}^{n}}$ is an admissible RPI set for the nominal closed-loop $\bar{A}+\bar{B}K_{t}$ in presence of disturbances $\mathbb{W}$ and constraints \eqref{eq.1.4} and \eqref{eq.1.3.1}.
\end{assum}
\begin{assum}\label{assum.stab2}
The linear feedback gain $K_t$ is such that $A_{K_{t}}=A+BK_{t}$ and $\bar{A}_{K_{t}}=\bar{A}+\bar{B}K_{t}$ are Schur.
\end{assum}

\begin{thm}[Stability]\label{thm.rstab}
Suppose Assumptions~\ref{assum.sets} to \ref{assum.stab2} hold. If (a) $Q$ is positive semidefinite and $R$ is positive definite, (b) $\bar{A}_{K}^{\top}P\bar{A}_{K}+\left(Q+K^{\top}RK\right)\leq P$, and (c) the nominal system is initialized such that $x(0)\in\left\{z(0)\right\}\oplus\mathbb{S}\subset\mathcal{Z}_{N}\oplus\mathbb{S}$, then the optimization \eqref{eq.1.9}--\eqref{eq.1.10} is feasible at all times, the state constraint \eqref{eq.1.4} and input constraint \eqref{eq.1.3.1} are met at all times and the set $\mathbb{A}\coloneqq\mathbb{S}\times\{\boldsymbol{0}\}$ is exponentially stable with a region of attraction $\left(\mathcal{Z}_N\oplus\mathbb{S}\right)\times\mathcal{Z}_N$ for the constrained composite closed-loop system
\begin{subequations}
\begin{alignat}{1}
	x(i+1)&=\bar{A}x(i)+\bar{B}\hat{\kappa}\left(x(i),z(i)\right)+w(i)\\
	z(i+1)&=\bar{A}z(i)+\bar{B}\kappa\left(z(i)\right).
\end{alignat}
\end{subequations}
\end{thm}

Assumption~\ref{assum.stab2} demands the knowledge of a linear feedback that stabilizes the unknown dynamics (which are possibly varying over time). This is a strong assumption; however, it is interesting to see that its realization is actually guaranteed by Assumption~\ref{assum.rpi}. First note that admissibility of $\mathbb{S}$ is necessary to guarantee that $\mathcal{Z}_{N}\neq\emptyset$, thus Assumption~\ref{assum.rpi} is required independently of Assumption~\ref{assum.stab2}. The following result establishes the link between both Assumptions.

\begin{prop}\label{prop.stab}
If Assumptions~\ref{assum.punc}, \ref{assum.stab} and \ref{assum.rpi} hold, then $A_{K_{t}}$ is Schur.
\end{prop}
\begin{pf}
Suppose $x(0)\in\mathbb{S}$ and so $z(i)=v(i)=\boldsymbol{0}$ for all $i$. Since $\mathbb{S}$ is constraint admissible, it follows that $w_{p}\in\mathbb{W}_{p}$, and assuming \eqref{eq.1.3.2} is met, then $w\in\mathbb{W}$. Therefore, for any $x\in\mathbb{S}$, it holds that $\bar{A}_{K_{t}}x+w\in\mathbb{S}$. It is easy to show that $\bar{A}_{K_{t}}x+w=A_{K_{t}}x+B\hat{w}$, and so, $A_{K_{t}}x+B\hat{w}\in\mathbb{S}$ for all $x\in\mathbb{S}$ and $\hat{w}\in\hat{\mathbb{W}}$. This implies that $\mathbb{S}$ is an RPI set for $A_{K_{t}}$ and disturbance $B\hat{\mathbb{W}}$, hence $A_{K_{t}}$ is Schur.\qed
\end{pf}
\begin{rem}
The admissibility of $\mathbb{S}$ depends on the size of $\mathbb{W}_{p}$, hence given $\left(\bar{A},\bar{B},\mathbb{X},\mathbb{U}\right)$ there is a bound on the parametric uncertainty that this approach can accept (i.e., a bound on $\mathcal{M}$).
\end{rem}

\section{Persistence of excitation}  \label{sec.pe}
The tube MPC controller is autonomous in the selection of the regulatory part of the input $\hat{u}(i)$, only requiring $\hat{w}(i)\in\hat{\mathbb{W}}$ for Theorem~\ref{thm.rstab} to hold. Consequently, the exciting sequence $\left\{\hat{w}(i)\right\}$ can be chosen independently to be of a certain PE order. The latter can be easily achieved in many different ways (e.g. with PRBS or sine signals \cite{Ljung1999}), however many such approaches lead to unnecessary perturbation affecting the plant. In order to design the exciting sequence in a way that considers its impact on the control objective, we propose the use of an MPC-like finite-horizon optimization problem, in which the decision variable is the exciting input. Define
\begin{equation} \label{eq.1.24}
\boldsymbol{\mathcal{M}}\left(\hat{w}(i)\right)=\sum_{j=0}^{l-1}\left(\boldsymbol{\hat{w}}_{i-j}\boldsymbol{\hat{w}}^{\top}_{i-j}\right)-\rho_{0}\mathcal{I}
\end{equation}
with $\boldsymbol{\hat{w}}_{i-j}$ as in \eqref{eq.1.8.2}. At each time instant the excitation $\hat{w}(i)$ to be applied to the system is obtained by solving
\begin{equation} \label{eq.1.15}
\hat{\mathbb{P}}_{h}\left(\left\{\hat{w}(i-1)\right\},x(i)\right):\min_{\hat{w}_{0} }\sum_{k=0}^{h-1}\ell\left(x_{k},\hat{w}_{k}\right)
\end{equation}
subject to (for $k=0,\ldots,h-1$):
\begin{subequations}  \label{eq.1.16}
\begin{alignat}{1}
x_{0}&=x(i)\\
x_{k+1}&=\bar{A}x_{k}+\bar{B}\hat{w}_{k}\\
\hat{w}_{0}&\in\hat{\mathbb{W}} \label{eq.1.16.1}\\
\hat{w}_{k+1}&=\hat{w}(i+k+1-l) \label{eq.1.16.2}\\
\boldsymbol{\mathcal{M}}\left(\hat{w}_{k}\right)&>0 \label{eq.1.16.3}.
\end{alignat}
\end{subequations}

The optimization \eqref{eq.1.15}--\eqref{eq.1.16} minimizes the running cost of the fictitious trajectories that would be generated by feeding the exciting part directly to the nominal model. Since $\hat{w}(i)$ is indeed a portion of the input, this helps minimize its disturbing impact on the plant. Albeit the prediction horizon in \eqref{eq.1.15}--\eqref{eq.1.16} is $h$, the solution is a single optimal value $\hat{w}^{*}_{0}(i)$ (or simply $\hat{w}^{*}_{0}$), while the remaining values are fixed by \eqref{eq.1.16.2} (similar to the idea of using different control and prediction horizons in nominal MPC). The reason for this is twofold, to reduce the complexity of the optimization, and to allow for a guarantee on recursive feasibility (in view of constraints \eqref{eq.1.16.1} and \eqref{eq.1.16.3}). Furthermore, \eqref{eq.1.15}--\eqref{eq.1.16} is implemented in a receding horizon fashion, so that feedback is introduced in the computation of $\hat{w}(i)$, to account for the parametric uncertainty.

\begin{rem}
Constraints \eqref{eq.1.16.2} and \eqref{eq.1.16.3} need access to the past values of $\hat{w}(i)$ over a time period of $l-1$ steps. This implies that a buffer sequence is required to initialize $\left\{\hat{w}(i)\right\}$ \cite{Marafioti2013,Hernandez2016}. Notice also that only the lower bound of \eqref{eq.1.8.1} is included in the definition of the PE measure \eqref{eq.1.24}, this is because the upper bound is trivially met given that $\hat{w}(i)$ is bounded \cite{Shouche1998}.
\end{rem}

\subsection{Recursive feasibility of the PE optimization}
The feasible space of \eqref{eq.1.15}--\eqref{eq.1.16} at time $i$ depends on past values of the exciting sequence, and the PE constraint \eqref{eq.1.16.3} is non-convex (see \cite{Marafioti2013}). However, despite the complexity of the problem, the structure of \eqref{eq.1.24} can be exploited to ensure that a feasible solution exists and it is known at each time instant, if there exist a buffer signal with certain characteristics.

\begin{assum} \label{assum.rfpe}
A buffer sequence $\left\{\hat{w}^{b}(h+l-2)\right\}$ is available and fulfils: (a) $\hat{w}^{b}(j)=\hat{w}^{b}(j-l)$ for all $j\geq l$, (b) $\hat{w}^{b}(j)\in\hat{\mathbb{W}}$ for all $j\in\left[0,h+l-2\right]$ and (c) $\boldsymbol{\mathcal{M}}(\hat{w}^{b}(h+l-2))>0$.
\end{assum}
\begin{prop} \label{prop.rfpe}
If Assumption~\ref{assum.rfpe} holds, and the exciting part is initialized as $\hat{w}(i)=\hat{w}^{b}(i)$ with $i\in\left[0,h+l-2\right]$, then for all $i\geq h+l-1$, $\hat{w}(i)=\hat{w}(i-l)$ is a feasible solution for \eqref{eq.1.15}--\eqref{eq.1.16}, and $\left\{\hat{w}(i)\right\}$ is SPE of order $h$.
\end{prop}

Proposition~\ref{prop.rfpe} guarantees that the periodic repetition of a particular buffer signal represents a feasible solution for \eqref{eq.1.15}--\eqref{eq.1.16}, however this sequence is computed off-line, and hence it is open-loop and not necessarily optimal. The following result provides a guarantee of recursive feasibility for exciting sequences generated in closed-loop by solving \eqref{eq.1.15}--\eqref{eq.1.16}.

\begin{thm}[RF of the PE optimization] \label{thm.rfpe}
If Assumption~\ref{assum.rfpe} holds, and the exciting part is initialized as $\hat{w}(i)=\hat{w}^{b}(i)$ with $i\in\left[0,h+l-2\right]$, then for all $i\geq h+l-1$ the optimization problem \eqref{eq.1.15}--\eqref{eq.1.16} is feasible, and $\left\{\hat{w}(i)\right\}$ is SPE of order $h$ with $\hat{w}(i)=\hat{w}^{*}_{0}$.
\end{thm}
\begin{pf}
Suppose that at time $i=h+l-1$ the optimizer has chosen $\hat{w}(i)=\hat{w}^{*}_{0}\neq\hat{w}(i-l)$ such that \eqref{eq.1.16} are fulfilled, then the new non-periodic value $\hat{w}(i)$ remains in $\hat{\boldsymbol{w}}_{j}$ for $h-1$ time steps. However, if $\hat{w}^{*}_{0}\neq\hat{w}(i-l)$ is a feasible solution at time $i$, then constraints \eqref{eq.1.16.2} and \eqref{eq.1.16.3} must have been satisfied at time $i$. This implies that periodic repetition of the past solution during $h-1$ future time steps remains feasible, and hence, $\hat{w}(j)=\hat{w}(j-l)$ is a feasible solution at time $j=i+1,\ldots,i+h-1$.
\end{pf}
\begin{rem} \label{rem.perio}
Given the non-convexity of \eqref{eq.1.16.3} and the time restrictions inherent to online optimization in a receding horizon framework, a minimum to \eqref{eq.1.15}--\eqref{eq.1.16} might not be found in time. However, Proposition~\ref{prop.rfpe} guarantees the existence of a solution at each time step thanks to constraint \eqref{eq.1.16.2}, which forces the new solution to satisfy Assumption~\ref{assum.rfpe}. Furthermore, Assumption~\ref{assum.rfpe} and constraint \eqref{eq.1.16.2}, as a way to guarantee the availability of a solution at each time instant, promote the generation of periodic exciting sequences (with period $l$).
\end{rem}

This approach to persistence of excitation has similarities with \cite{Marafioti2013}. The key difference is that in \cite{Marafioti2013} the whole input is used to excite the system, and hence stability is only an assumption. Another significant difference is that \cite{Marafioti2013} employs a single constraint, $\boldsymbol{\mathcal{M}}\left(\hat{w}_{0}\right)>0$, to achieve the required SPE behaviour. This is evidently less demanding than \eqref{eq.1.16}, but it also yields a weaker result. A similar claim to that in Proposition~\ref{prop.rfpe} is provided, however there is no guidance as to how the buffer signal should be designed, as opposed to the structure described in Assumption~\ref{assum.rfpe}. Furthermore, there is no feasibility guarantees after a time step in which the optimizer sets $\hat{w}(i)=\hat{w}^{*}_{0}\neq\hat{w}(i-l)$, contrary to the result provided by Theorem~\ref{thm.rfpe}.

\subsection{Transmissibility of the persistence of excitation}
Theorem~\ref{thm.rfpe} guarantees that a solution to \eqref{eq.1.15}--\eqref{eq.1.16} exists and that it results in an exciting sequence that is SPE of order $h$. However, setting $h=n+m$ is not sufficient to meet the requirements of Corollary~\ref{cor.pe1}. This is because the robust control policy \eqref{eq.1.11} is designed precisely to reject the whole disturbance $w(i)=B\hat{w}(i)+w_{p}(i)$, thereby the true input sequence $\left\{u(i)\right\}$ might not inherit the SPE order.

Transmissibility of the SPE order from $\left\{\hat{w}(i)\right\}$ to $\left\{u(i)\right\}$, and hence to the regressor, can be achieved under design requirements that are marginally more demanding than those of Corollary~\ref{cor.pe1}. The whole input fed to the plant is
\begin{equation}\label{eq.1.40}
u(i)=\left(v(i)-K_{t}z(i)\right)+\left(K_{t}x(i)+\hat{w}(i)\right).
\end{equation}
Given Theorem~\ref{thm.rstab} and Lemma 2.2 from \cite{Bai1985}, the excitation properties of the plant input depend solely on $K_{t}x(i)+\hat{w}(i)$. By neglecting the converging term in \eqref{eq.1.40}, the following state space model can be constructed
\begin{subequations} \label{eq.1.14}
\begin{alignat}{2}
x(i+1)&=\left(A+BK_{t}\right)x(i)+B\hat{w}(i)\\
u(i)&=K_{t}x(i)+\hat{w}(i),
\end{alignat}
\end{subequations}
with input $\hat{w}(i)$ and output $u(i)$. In view of Lemma~\ref{lem.pe1}, the matter of transmissibility of the PE condition from $\left\{\hat{w}(i)\right\}$ to $\left\{u(i)\right\}$, reduces to the question of reachability of $u(i)$ from $\hat{w}(i)$.
\begin{defn}[Output Reachability]\label{defin.oreach}
Consider a state-space system with input $u(i)\in\mathbb{R}^{m}$, output $y(i)\in\mathbb{R}^{p}$ and state $x(i)\in\mathbb{R}^{n}$
\begin{subequations} \label{eq.1.26}
\begin{alignat}{1}
x(i+1)&=Ax(i)+Bu(i)\\
y(i)&=Cx(i)+Du(i),
\end{alignat}
\end{subequations}
where $A$, $B$, $C$ and $D$ are of appropriate dimension. System \eqref{eq.1.26} is said to be output reachable if $\mathcal{O}_{o}=\left[D\:\:CB\:\:CAB\:\:\cdots\:\:CA^{n-1}B\right]$ has full row rank.
\end{defn}

\begin{thm}[Transmissibility of PE condition] \label{thm.petrans}
If the origin is asymptotically stable for the undisturbed nominal closed-loop system, and the sequence $\left\{\hat{w}(i)\right\}$ is SPE of order $2n+m$, then the regressor vector sequence $\left\{\phi(i)\right\}$ is SPE of order 1.
\end{thm}
\begin{pf}
The output reachability matrix for system \eqref{eq.1.14} is
\begin{equation}
\mathcal{O}_{o}=\left[\mathcal{I}\:\:K_{t}B\:\:K_{t}A^{\vphantom{n-1}}_{K_{t}}B\:\:\cdots\:\:K_{t}A^{n-1}_{K_{t}}B\right].
\end{equation}
Since $\mathcal{I}$ has row rank $m$, $\mathcal{O}_{o}$ has full row rank, and $u(i)$ is reachable from $\hat{w}(i)$. The McMillan degree of system \eqref{eq.1.14} is $n$, and the exciting sequence $\left\{\hat{w}(i)\right\}$ is SPE of degree $2n+m$, thus Lemma~\ref{lem.pe1} ensures that the input sequence $\left\{u(i)\right\}$ is SPE of degree $n+m$. According to Corollary~\ref{cor.pe1}, this guarantees that the corresponding regressor sequence $\left\{\phi(i)\right\}$ is SPE of order 1.
\qed
\end{pf}

\section{Prediction model update} \label{sec.pmu}
Given the deterministic framework, Theorem~\ref{thm.petrans} guarantees that the estimates in \eqref{eq.1.6.3} will converge to the true values of the plant parameters. Since the closed-loop performance of the robust controller depends on the accuracy of the prediction model, it would be desirable to use the converged estimates to replace the model in \eqref{eq.1.10.1}; however, this might not be possible, because the stability and admissibility guarantees depend on parameters designed specifically for $\bigl(\bar{A},\bar{B}\bigr)$.

This appears as a shortcoming when compared to the continuous model update allowed by other approaches such as \cite{Adetola2009,Fukushima2007,Fan2016}, ensured by the assumptions and formulations employed therein. However, this continuous adaptation has a direct impact in the complexity of the resulting controllers, for example requiring the online re-computation of various elements at each time instant in \cite{Adetola2009,Fukushima2007}. The adaptive control approach proposed in this paper is simpler in formulation and design, but at the price of not necessarily being able to update the prediction model with the true plant parameters.

Various approaches can be employed to verify and/or ensure that a certain set of estimates can be used as a new prediction model for the AMPC controller devised in this paper. Our aim is not to propose a particular solution, but to present a range of options that vary between no adjustment to the controller, to its full re-design; the most suitable will depend on the particular application. Moreover, we aim to highlight some of the issues surrounding AMPC with simultaneous control and excitation guarantees. The options we present aim to maintain the robustness properties of the controller in order to account for LTV systems, i.e., cases in which the plant may change again after parameter convergence (although remaining inside $\mathcal{M}$).

\subsection{Simple a-posteriori verification} \label{sec.verification}
The most straightforward approach is to simply verify whether the necessary conditions for Theorem~\ref{thm.rstab} to hold are still met if all the controller parameters remain fixed but the prediction model is updated by a set of converged estimates represented by $\bigl(\tilde{A},\tilde{B}\bigr)$. In particular, the following set of conditions would have to be verified if a model update is to be performed at time $i$:
\begin{enumerate}[(a)]
\item \textbf{Assumption~\ref{assum.punc}}, there exists a $\mathcal{C}$-set $\tilde{\mathbb{W}}_{p}$ such that, $w_{p}\in\tilde{\mathbb{W}}_{p}$ for all $\left(x,u,\left[A\:\:B\right]\right)\in\mathbb{X}\times\mathbb{U}\times\mathcal{M}$ and nominal model $\bigl(\tilde{A},\tilde{B}\bigr)$. \label{condition.1}
\item \textbf{Assumption~\ref{assum.pi}}, the set $\mathbb{Z}_{f}$ is a PI set for the closed-loop dynamics $\tilde{A}+\tilde{B}K$. \label{condition.2}
\item \textbf{Assumption~\ref{assum.rpi}}, the set $\mathbb{S}$ is an RPI set for the closed-loop dynamics $\tilde{A}+\tilde{B}K_{t}$ and disturbance $\tilde{\mathbb{W}}_{p}$.  \label{condition.3}
\item \textbf{Feasibility}, there exists a feasible solution to \eqref{eq.1.9}--\eqref{eq.1.10} with \eqref{eq.1.10.1} replaced by $z_{k+1}=\tilde{A}z_{k}+\tilde{B}v_{k}$ at time $i$. Furthermore $\min_{\boldsymbol{v}}V_{N}\left(z(i),\boldsymbol{v},\tilde{A},\tilde{B}\right)-\min_{\boldsymbol{v}}V_{N}\left(z(i-1),\boldsymbol{v},\bar{A},\bar{B}\right)<0.$ \label{condition.4}
\item \textbf{Stability}, for $\tilde{A}_{K}=\tilde{A}+\tilde{B}K$, it holds that
\begin{equation*}
\tilde{A}_{K}^{\top}P\tilde{A}_{K}+\left(Q+K^{\top}RK\right)\leq P.
\end{equation*}  \label{condition.5}
\end{enumerate}
If conditions \eqref{condition.1}--\eqref{condition.5} are met, then Theorem~\ref{thm.rstab} holds under an update of the prediction model. It is easy to show that there exists a finite time instant for which \eqref{condition.4} is met irrespective of $\bigl(\tilde{A},\tilde{B}\bigr)$, however it is also easy to find an example for which \eqref{condition.1} is not met by any $\left[A\:\:B\right]\in\mathcal{M}$ and thus the prediction model could never be updated.

\subsection{Complete controller redesign}
A completely opposite approach is to recompute the entirety of the controller parameters online, and wait for an appropriate time instant $i$ in which admissibility and stability conditions allow for a complete update.
\begin{prop}\label{prop.mup1}
Assume that, for a certain pair $\bigl(\tilde{A},\tilde{B}\bigr)$, a set of parameters $\left(\tilde{\mathbb{W}}_{p},\tilde{K}_{t},\tilde{\mathbb{S}},\tilde{K},\tilde{\mathbb{Z}}_{f},\tilde{P}\right)$ that fulfils Assumptions~\ref{assum.punc}, \ref{assum.pi}, \ref{assum.rpi} and (b) from Theorem~\ref{thm.rstab} is available. Define the associated optimization problem as $\tilde{\mathbb{P}}_{N}\left(z=z(i)\right)$ with cost $\tilde{V}_{N}\left(z(i),\boldsymbol{v}\right)$. Then Theorem~\ref{thm.rstab} holds under a prediction model update performed at any time instant $i$ for which $\tilde{\mathbb{P}}_{N}\left(z=z(i)\right)$ is feasible and $\min_{\boldsymbol{v}}V_{N}\left(z(i),\boldsymbol{v},\tilde{A},\tilde{B}\right)-\min_{\boldsymbol{v}}V_{N}\left(z(i-1),\boldsymbol{v},\bar{A},\bar{B}\right)<0$.
\end{prop}
Note that, similarly to the previous case, it is straightforward to show that there exists a finite time instant for which the hypotheses of Proposition~\ref{prop.mup1} are met irrespective of $\bigl(\tilde{A},\tilde{B}\bigr)$ (except for the admissibility of the RPI set $\mathbb{S}$).

A drawback of this approach is that the computational cost of computing invariant sets grows rapidly with the size of the plant, thus making it difficult to compute them online in most applications. However, note that albeit desirable, the prediction model need not to be updated instantaneously after the estimates have converged, thus the re-computation can be performed during multiple sampling periods. Furthermore there exists efficient methods to compute invariant approximations to the type of set usually employed in MPC implementations, such as the minimal RPI set (see for example \cite{Rakovic2005,Trodden2016a}), hence making this approach a feasible solution.

\subsection{Robustification of the initial design} \label{sec.verification2}
A third approach is to robustify the design procedure in order to increase the chances of a certain subset of models $\tilde{\mathcal{M}}\subset\mathcal{M}$ to contain a feasible replacement for the prediction model under minimal controller redesign. A comprehensive robustification of the entire design procedure would contradict the simplicity that is a key feature of the proposed approach. Instead, consider the following assumption.

\begin{assum} \label{assum.mup}
The set $\mathbb{Z}_{f}$ is a $\lambda$-contractive set \cite{Li1994} for the closed-loop dynamics $\bar{A}+\bar{B}K$ such that $\left(1-\lambda\right)\mathbb{Z}_{f}\supseteq\mathbb{W}_{f}$ with $\mathbb{W}_{f}$ a $\mathcal{C}$-set that contains the parametric uncertainty for all $\left(z,\left[A\:\:B\right]\right)\in\mathbb{Z}_{f}\times\tilde{\mathcal{M}}$ and nominal model $\bigl(\bar{A},\bar{B}\bigr)$. Furthermore, there exists a matrix $\tilde{P}$ that fulfils $\tilde{A}_{K}^{\top}\tilde{P}\tilde{A}_{K}+\left(Q+K^{\top}RK\right)\leq \tilde{P}$.
\end{assum}

Clearly then, for a particular $\bigl[\tilde{A}\:\tilde{B}\bigr]^{\top}\in\tilde{\mathcal{M}}$ to be a feasible prediction model, the following conditions have to be verified:
\begin{enumerate}[(a)]
\item \textbf{Assumption~\ref{assum.punc}}, there exists a $\mathcal{C}$-set $\tilde{\mathbb{W}}_{p}$ such that, $w_{p}\in\tilde{\mathbb{W}}_{p}$ for all $\left(x,u,\left[A\:\:B\right]\right)\in\mathbb{X}\times\mathbb{U}\times\mathcal{M}$ and nominal model $\bigl(\tilde{A},\tilde{B}\bigr)$. \label{condition.6}
\item \textbf{Assumption~\ref{assum.rpi}}, the set $\mathbb{S}$ is an RPI set for the closed-loop dynamics $\tilde{A}+\tilde{B}K_{t}$ and disturbance $\tilde{\mathbb{W}}_{p}$.  \label{condition.7}
\item \textbf{Feasibility}, there exists a feasible solution to \eqref{eq.1.9}--\eqref{eq.1.10} with \eqref{eq.1.10.1} replaced by $z_{k+1}=\tilde{A}z_{k}+\tilde{B}v_{k}$ at time $i$. Furthermore $\min_{\boldsymbol{v}}V_{N}\left(z(i),\boldsymbol{v},\tilde{A},\tilde{B}\right)-\min_{\boldsymbol{v}}V_{N}\left(z(i-1),\boldsymbol{v},\bar{A},\bar{B}\right)<0.$\label{condition.8}
\end{enumerate}
Similarly to the previous approaches, it is possible to show that there exists a finite time for which condition \eqref{condition.8} is met. It is evident then that by meeting Assumption~\ref{assum.mup}, the set of conditions that must be verified a-posteriori is less restrictive, thus increasing the chance of a pair $\bigl[\tilde{A}\:\:\tilde{B}\bigr]^{\top}\in\tilde{\mathcal{M}}$ to be feasible for model update. Moreover, the computation of $\lambda$-contractive sets is as complex as that of PI sets, and the search of a $\lambda$ to meet Assumption~\ref{assum.mup} is performed off-line, and can be done iteratively by decreasing the value of $\lambda$ until the inclusion  $\mathbb{W}_{f}\subseteq\left(1-\lambda\right)\mathbb{Z}_{f}$ is verified. Consequently, Assumption~\ref{assum.mup} does not increase the complexity of the design procedure.

\section{Illustrative example} \label{sec.numerical}
Consider a point-mass spring-damper plant (representing a truck), where the control objective is to steer the mass to an arbitrary horizontal equilibrium using an horizontally applied force $u$. The dynamics of the plant are described by
\begin{equation} \label{eq.1.30}
\dot{x}=\begin{bmatrix}0& 1\\\nicefrac{-c}{M}&\nicefrac{-k}{M}\end{bmatrix}x+\begin{bmatrix}0\\\nicefrac{100}{M}\end{bmatrix}u,
\end{equation}
where the state vector is composed by the horizontal position and velocity of the truck. The plant is subject to bound constraints on the states $||x||_{\infty}\leq 15$ and inputs $|u|\leq 5$, thus Assumption~\ref{assum.sets} is met. Moreover, for any $M>0$, the (continuous time) system is state reachable, and so Lemma~\ref{lem.pe1} is applicable. A sampling time $T_{s}=0.1[s]$ is used to discretize the plant.

Initially, the truck is loaded with a mass of 2[Kg], and the spring and damper coefficients are assumed to be at factory values of 10[\nicefrac{N}{m}] and 30[\nicefrac{N}{ms}] respectively. During operation, increasing temperatures may result in the spring losing up to 25\% of its stiffness; furthermore, an uncontrolled external disturbance may increase the truck's load by 25\% at an arbitrary time. This information results in a compact (non-convex) set $\mathcal{M}$ in which the true plant is expected to reside at all times. The initial conditions of the mass, spring and damper are used to define the nominal prediction model $\bigl(\bar{A},\bar{B}\bigr)$. In view of this, a set $\mathbb{W}_{p}$ that fulfils Assumption~\ref{assum.punc} can be computed as the convex hull of the individual uncertainty sets arising from each different plant configuration. 

The PE related parameters are set to $\alpha=0.9$ and $\rho=0.05$. A horizon of $N=3$ is employed for the nominal MPC. Larger horizons would increase the size of $\mathcal{Z}_{N}$, however $\mathbb{S}$, the main source of conservatism of the proposed approach, would remain unchanged. The weight matrices are set to $Q=\text{diag}\left(100,1\right)$ and $R=1$, and the linear gains $K=K_{t}$ and terminal cost $P$ are computed as the corresponding infinite horizon optimal linear gain and cost, hence meeting Assumption~\ref{assum.stab}. The sets $\mathbb{S}$ and $\mathbb{Z}_{f}$ are computed as the minimal RPI set and the maximal PI set for the closed loop $\bar{A}+\bar{B}K$ and disturbance $\mathbb{W}$, resulting in Assumptions~\ref{assum.pi}--\ref{assum.stab2} being met. A buffer sequence for the exciting part is computed following Assumption~\ref{assum.rfpe}. The SPE order is set to $h=5$, and $l=h$ (given the values of $\rho_{0}$ and $\alpha$). Finally, the forgetting factor for the RLS algorithm is set to $\lambda=0.75$.

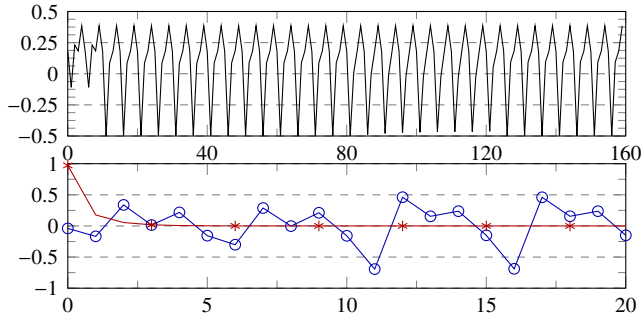
\begin{figure}
\begin{center}
\setlength\figureheight{0.2\textwidth}
\setlength\figurewidth{0.4\textwidth}
\pgfplotsset{major grid style={dashed,gray}}
\pgfplotsset{minor grid style={dotted,gray}}
\newcommand{\picheight}{0.45}
\newcommand{\picwidth}{0.45}
\newcommand{\picwidthh}{0.55}

\begin{tikzpicture}
[align=center,
square node/.style={rectangle,draw=black,fill=none,thin,minimum width=2.5mm,minimum height=2.5mm},
every pin/.style={fill=white}]

\begin{axis}[%
width=\figurewidth,
height=\picheight\figureheight,
at={(0\figurewidth,1\figureheight)},
scale only axis,
xmin=0,
xmax=160,
xtick={0,40,...,160},
ymin=-0.5,
ymax=0.5,
ytick={-0.5,-0.25,...,0.5},
minor x tick num={1},
minor y tick num={3},
xticklabel style = {font=\footnotesize},
yticklabel style = {font=\footnotesize},
ymajorgrids,
%yminorgrids,
%xmajorgrids,
%xminorgrids,
]
\addplot [color=black,solid,forget plot]
  table[row sep=crcr]{%
0	0.180023718797403\\
1	-0.108941213814203\\
2	0.230477856774093\\
3	0.183362424454117\\
4	0.384485844350483\\
5	0.180023718797403\\
6	-0.108941213814203\\
7	0.230477856774093\\
8	0.183362424454117\\
9	0.384485844350483\\
10	0.180023718797403\\
11	-0.499994782339045\\
12	0.086368978155194\\
13	0.183362424454117\\
14	0.384485844350483\\
15	0.180023718797403\\
16	-0.49793994511232\\
17	0.0855001830981402\\
18	0.183362424454117\\
19	0.384485844350483\\
20	0.180023718797403\\
21	-0.497275661494742\\
22	0.0848417414719719\\
23	0.183362424454117\\
24	0.384485844350483\\
25	0.180023718797403\\
26	-0.496800652565417\\
27	0.0843681725091251\\
28	0.183362424454117\\
29	0.384485844350483\\
30	0.180023718797403\\
31	-0.496457856933689\\
32	0.0840264004498254\\
33	0.183362424454117\\
34	0.384485844350483\\
35	0.180023718797403\\
36	-0.496209959149238\\
37	0.0837792417505276\\
38	0.183362424454117\\
39	0.384485844350483\\
40	0.180023718797403\\
41	-0.499986859002266\\
42	0.0869882953793669\\
43	0.183362424454117\\
44	0.384485844350483\\
45	0.180023718797403\\
46	-0.499991491088135\\
47	0.0876125752406287\\
48	0.183362424454117\\
49	0.384485844350483\\
50	0.180023718797403\\
51	-0.499994054315736\\
52	0.0877685500661169\\
53	0.183362424454117\\
54	0.384485844350483\\
55	0.180023718797403\\
56	-0.499994541126962\\
57	0.0878056132678291\\
58	0.183362424454117\\
59	0.384485844350483\\
60	0.180023718797403\\
61	-0.499994648219211\\
62	0.0878181341573648\\
63	0.183362424454117\\
64	0.384485844350483\\
65	0.180023718797403\\
66	-0.499994679807085\\
67	0.0878164954100661\\
68	0.183362424454117\\
69	0.384485844350483\\
70	0.180023718797403\\
71	-0.499994679085616\\
72	0.087817002602216\\
73	0.183362424454117\\
74	0.384485844350483\\
75	0.180023718797403\\
76	-0.499994680488883\\
77	0.0878171198257306\\
78	0.183362424454117\\
79	0.384485844350483\\
80	0.180023718797403\\
81	-0.499999953411222\\
82	0.00896393110316839\\
83	0.183362424454117\\
84	0.384485844350483\\
85	0.180023718797403\\
86	-0.487946688293909\\
87	-0.00302895530162242\\
88	0.183362424454117\\
89	0.384485844350483\\
90	0.180023718797403\\
91	-0.479252182283407\\
92	-0.0109519697163372\\
93	0.183362424454117\\
94	0.384485844350483\\
95	0.180023718797403\\
96	-0.473497184233974\\
97	-0.0161972681173777\\
98	0.183362424454117\\
99	0.384485844350483\\
100	0.180023718797403\\
101	-0.46968837430746\\
102	-0.0196687475727409\\
103	0.183362424454117\\
104	0.384485844350483\\
105	0.180023718797403\\
106	-0.467166631243733\\
107	-0.0219671455948003\\
108	0.183362424454117\\
109	0.384485844350483\\
110	0.180023718797403\\
111	-0.465497031371147\\
112	-0.0234888719067094\\
113	0.183362424454117\\
114	0.384485844350483\\
115	0.180023718797403\\
116	-0.464391622278482\\
117	-0.024496385306782\\
118	0.183362424454117\\
119	0.384485844350483\\
120	0.180023718797403\\
121	-0.462133060685099\\
122	0.054584142536253\\
123	0.183362424454117\\
124	0.384485844350483\\
125	0.180023718797403\\
126	-0.499999850758529\\
127	0.093912603363076\\
128	0.183362424454117\\
129	0.384485844350483\\
130	0.180023718797403\\
131	-0.499999948646002\\
132	0.0939719524056906\\
133	0.183362424454117\\
134	0.384485844350483\\
135	0.180023718797403\\
136	-0.499999948702674\\
137	0.0939704880514377\\
138	0.183362424454117\\
139	0.384485844350483\\
140	0.180023718797403\\
141	-0.499999948701422\\
142	0.0939704880514377\\
143	0.183362424454117\\
144	0.384485844350483\\
145	0.180023718797403\\
146	-0.499999948701561\\
147	0.0939704880514377\\
148	0.183362424454117\\
149	0.384485844350483\\
150	0.180023718797403\\
151	-0.499999948701881\\
152	0.0939704880514377\\
153	0.183362424454117\\
154	0.384485844350483\\
155	0.180023718797403\\
156	-0.49999994870144\\
157	0.0939704880514377\\
158	0.183362424454117\\
159	0.384485844350483\\
};\label{trajectory.wh}
\coordinate (pt1) at (axis cs:9,0);
\coordinate (pt2) at (axis cs:100,0);
\end{axis}
\begin{axis}[%
width=\figurewidth,
height=\picheight\figureheight,
at={(0\figurewidth,\picheight\figureheight)},
scale only axis,
xmin=0,
xmax=20,
xtick={0,5,...,20},
ymin=-1,
ymax=1,
ytick={-1,-0.5,...,1},
minor x tick num={1},
minor y tick num={1},
xticklabel style = {font=\footnotesize},
yticklabel style = {font=\footnotesize},
ymajorgrids,
%yminorgrids,
%xmajorgrids,
%xminorgrids,
]
\addplot [color=blue!70!black,mark=o,mark repeat={1},solid,forget plot]
table[row sep=crcr]{%
0	-0.0399210489613721\\
1	-0.167264701466851\\
2	0.337562056443129\\
3	0.0145968853118694\\
4	0.217456860835119\\
5	-0.156339150211539\\
6	-0.301470123704349\\
7	0.287718115644164\\
8	-0.00270288027761612\\
9	0.211495980711464\\
10	-0.158391370102243\\
11	-0.693230168301551\\
12	0.461831002782238\\
13	0.154488160524589\\
14	0.237218434710545\\
15	-0.15063555518621\\
16	-0.688547960671759\\
17	0.46019155282903\\
18	0.155296477252472\\
19	0.237356208306569\\
20	-0.15059357415449\\
21	-0.687869436055533\\
22	0.458997028387767\\
23	0.155766576477142\\
24	0.237405239726831\\
25	-0.150581062079606\\
26	-0.687390288929586\\
27	0.458138041252229\\
28	0.156104242747288\\
29	0.237440333308349\\
30	-0.150572122709775\\
31	-0.687044537558839\\
32	0.457518117224136\\
33	0.156347932552514\\
34	0.237465659337185\\
35	-0.15056567150042\\
36	-0.686794506731945\\
37	0.457069807248333\\
38	0.156524160870702\\
39	0.237483974168056\\
40	-0.150561006242245\\
41	-0.692645641537909\\
42	0.462082494411328\\
43	0.155243716018069\\
44	0.237559900265068\\
45	-0.151655388227499\\
46	-0.692790782670862\\
47	0.462669912306081\\
48	0.154721048068616\\
49	0.237488858751165\\
50	-0.151676325127513\\
51	-0.692800570393786\\
52	0.462825439413559\\
53	0.154593211857575\\
54	0.237472065776935\\
55	-0.151681219802352\\
56	-0.692802743817395\\
57	0.462862307161683\\
58	0.154562825812857\\
59	0.237468072931167\\
60	-0.151682383723158\\
61	-0.692803251979983\\
62	0.462874774501758\\
63	0.154552575408864\\
64	0.237466729622089\\
65	-0.151682774948645\\
66	-0.692803418363241\\
67	0.462873114165132\\
68	0.154553898674397\\
69	0.237466898688737\\
70	-0.151682726130867\\
71	-0.692803400840133\\
72	0.462873626670475\\
73	0.154553487003602\\
74	0.237466845557041\\
75	-0.151682741525718\\
76	-0.692803407544187\\
77	0.462873743176505\\
78	0.154553390888337\\
79	0.237466832925742\\
80	-0.151682745207895\\
81	-0.728568413993047\\
82	0.314468719760606\\
83	0.288357247111196\\
84	0.274820932398113\\
85	-0.139676314808594\\
86	-0.715834690532957\\
87	0.29266846442694\\
88	0.29355335507017\\
89	0.276833171790327\\
90	-0.139190155240345\\
91	-0.707116620964071\\
92	0.278422527644946\\
93	0.297256367849506\\
94	0.278295021612413\\
95	-0.138828056845611\\
96	-0.701339495476029\\
97	0.268990680737243\\
98	0.299706571765082\\
99	0.279262146615975\\
100	-0.138588548183838\\
101	-0.697516072978806\\
102	0.262748431865866\\
103	0.301328186606944\\
104	0.279902218093177\\
105	-0.13843003420315\\
106	-0.694984658790989\\
107	0.25861556379101\\
108	0.302401828093359\\
109	0.280325997872457\\
110	-0.138325084900119\\
111	-0.693308655810276\\
112	0.255879268479665\\
113	0.303112665796385\\
114	0.28060657449168\\
115	-0.138255599925315\\
116	-0.692199007331365\\
117	0.254067612933029\\
118	0.303583303506102\\
119	0.280792340661758\\
120	-0.138209594967655\\
121	-0.688224911296715\\
122	0.333232318336059\\
123	0.245730689377592\\
124	0.261988188186639\\
125	-0.140766727307723\\
126	-0.725295055847123\\
127	0.400622017822744\\
128	0.226219701705806\\
129	0.254574497950602\\
130	-0.142467267300696\\
131	-0.725319380010958\\
132	0.400879672900077\\
133	0.226290381453949\\
134	0.254599665747502\\
135	-0.142462047712616\\
136	-0.725319606075147\\
137	0.400877432808358\\
138	0.22629103314558\\
139	0.254599866628031\\
140	-0.142462016952481\\
141	-0.725319613950146\\
142	0.400877424590061\\
143	0.226291029548408\\
144	0.254599865647434\\
145	-0.142462017050874\\
146	-0.725319613883028\\
147	0.400877424639853\\
148	0.226291029567722\\
149	0.254599865651971\\
150	-0.142462017050774\\
151	-0.725319613883879\\
152	0.400877424639754\\
153	0.226291029567656\\
154	0.254599865651927\\
155	-0.14246201705081\\
156	-0.72531961388347\\
157	0.400877424639405\\
158	0.226291029567526\\
159	0.254599865651881\\
};
\label{trajectory.u}
\addplot [color=red!70!black,solid,mark=asterisk,mark repeat={3},forget plot]
table[row sep=crcr]{%
0	0.972243278749222\\
1	0.176559802937867\\
2	0.0546583381094574\\
3	0.0185790250313601\\
4	0.00638663128259514\\
5	0.00219821770026621\\
6	0.000756716393844328\\
7	0.000260496939533779\\
8	8.96681728293931e-05\\
9	3.08737544703099e-05\\
10	1.06287419309716e-05\\
11	3.66174132016785e-06\\
12	1.32341233489243e-06\\
13	6.32091531836736e-07\\
14	0\\
15	0\\
16	0\\
17	0\\
18	0\\
19	0\\
20	0\\
21	0\\
22	0\\
23	0\\
24	0\\
25	0\\
26	0\\
27	0\\
28	0\\
29	0\\
30	0\\
31	0\\
32	0\\
33	0\\
34	0\\
35	0\\
36	0\\
37	0\\
38	0\\
39	0\\
40	0\\
41	0\\
42	0\\
43	0\\
44	0\\
45	0\\
46	0\\
47	0\\
48	0\\
49	0\\
50	0\\
51	0\\
52	0\\
53	0\\
54	0\\
55	0\\
56	0\\
57	0\\
58	0\\
59	0\\
60	0\\
61	0\\
62	0\\
63	0\\
64	0\\
65	0\\
66	0\\
67	0\\
68	0\\
69	0\\
70	0\\
71	0\\
72	0\\
73	0\\
74	0\\
75	0\\
76	0\\
77	0\\
78	0\\
79	0\\
80	0\\
81	0\\
82	0\\
83	0\\
84	0\\
85	0\\
86	0\\
87	0\\
88	0\\
89	0\\
90	0\\
91	0\\
92	0\\
93	0\\
94	0\\
95	0\\
96	0\\
97	0\\
98	0\\
99	0\\
100	0\\
101	0\\
102	0\\
103	0\\
104	0\\
105	0\\
106	0\\
107	0\\
108	0\\
109	0\\
110	0\\
111	0\\
112	0\\
113	0\\
114	0\\
115	0\\
116	0\\
117	0\\
118	0\\
119	0\\
120	0\\
121	0\\
122	0\\
123	0\\
124	0\\
125	0\\
126	0\\
127	0\\
128	0\\
129	0\\
130	0\\
131	0\\
132	0\\
133	0\\
134	0\\
135	0\\
136	0\\
137	0\\
138	0\\
139	0\\
140	0\\
141	0\\
142	0\\
143	0\\
144	0\\
145	0\\
146	0\\
147	0\\
148	0\\
149	0\\
150	0\\
151	0\\
152	0\\
153	0\\
154	0\\
155	0\\
156	0\\
157	0\\
158	0\\
159	0\\
};\label{trajectory.v}
\coordinate (pt3) at (axis cs:10,1.5);
\end{axis}
%\node[square node, name=z1, minimum height=2cm,minimum width=0.7cm,line width=0.4pt] at (pt1) {};
%\node[square node, name=z2, minimum height=2cm,minimum width=0.7cm,line width=0.4pt] at (pt2) {};
\end{tikzpicture}%
\vspace*{-0.4cm}
\caption{(Top) Exciting sequence \ref{trajectory.wh} $\hat{w}(i)$, (Bottom) Input sequences \ref{trajectory.u} $u(i)$, \ref{trajectory.v} $v(i)$.} 
\label{fig.inputs}
\end{center}
\end{figure}

The system is initialized at $x(0)=[-1\:\:15]^{\top}$ and it changes between different operating conditions as follows: (a) nominal, (b) 25\% stiffness loss at $i=40$, (c) 25\% load increase at $i=80$, (d) stiffness restoration at $i=120$. Figure~\ref{fig.inputs} shows the optimized exciting sequence $\left\{\hat{w}(i)\right\}$. At time instant $i=9$ the optimizer takes over the buffer and sets $\hat{w}^{*}_{0}(i)\neq\hat{w}^{b}(i-l)$, however due to the non-convexity of the optimization and the fact that operating condition (b) is a feasible prediction model, the sequence remains periodic until the plant changes into mode (c). Similarly, the periodicity observed after $i=100$ is broken once a new set of estimates, closer to mode (d), is employed for predictions. Figure~\ref{fig.inputs} also shows that the nominal input sequence $\left\{v(i)\right\}$ converges after 15 time steps, but the true input sequence $\left\{u(i)\right\}$ remains disturbed thanks to the action of the exciting part.

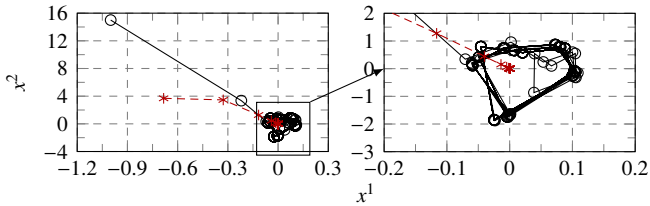
\begin{figure}
\begin{center}
\setlength\figureheight{0.1\textwidth}
\setlength\figurewidth{0.4\textwidth}
\pgfplotsset{major grid style={dashed,gray}}
\pgfplotsset{minor grid style={dotted,gray}}
\newcommand{\picwidth}{0.45}
\newcommand{\picwidthh}{0.55}

\begin{tikzpicture}[align=center,
square node/.style={rectangle,draw=black,fill=none,thin,minimum width=2.5mm,minimum height=2.5mm},
every pin/.style={fill=white}]
\begin{axis}[%
width=\picwidth\figurewidth,
height=\figureheight,
at={(0\figurewidth,1\figureheight)},
scale only axis,
xmin=-1.2,
xmax=0.3,
xtick={-1.2,-0.9,-0.6,-0.3,0,0.3},
ymin=-4,
ymax=16,
ytick={-4,0,...,16},
minor x tick num={1},
minor y tick num={1},
yticklabel style = {font=\footnotesize},
xticklabel style = {font=\footnotesize},
ylabel={\footnotesize$x^{2}$},
xlabel={\footnotesize$x^{1}$},
x label style={at={(3.8cm,-0.3cm)}},
xmajorgrids,
ymajorgrids,
tick label style={/pgf/number format/fixed},
]
\addplot [color=black,solid,mark=o,mark options={solid},forget plot]
  table[row sep=crcr]{%
-1	15\\
-0.219739665635995	3.35455608422688\\
-0.0706565690469134	0.342465256627355\\
0.00211266487154647	0.958346046037598\\
0.0536518911113192	0.241415387649937\\
0.100010630998353	0.596327877935451\\
0.104016667947759	-0.300070142439352\\
0.0386697347797844	-0.8651575159184\\
0.0396657077396008	0.544578539695769\\
0.0665762622653818	0.0990523969273757\\
0.104459692800299	0.547322608693661\\
0.105548240936322	-0.316939966563813\\
-0.0234104773060279	-1.87550555610258\\
-0.0454529876027668	0.792231862493189\\
0.0207099735690577	0.577547410335864\\
0.0880290858580497	0.727266866762162\\
0.099867225534307	-0.254405274040538\\
-0.0250381316483922	-1.84867685952447\\
-0.0459387040830676	0.794162404364296\\
0.0204606281683812	0.580160468015114\\
0.0879400386892054	0.728242323848081\\
0.0998364403403215	-0.254066446297423\\
-0.0249423849177627	-1.84685366512932\\
-0.0459420642852919	0.791458283097746\\
0.0203936571334545	0.580792005030567\\
0.0879144356774874	0.728520216967713\\
0.0998275274763142	-0.253968440258361\\
-0.0248694087374712	-1.84559963026487\\
-0.0459433523579183	0.78950051207318\\
0.0203458671893661	0.581242047796113\\
0.0878961507589464	0.728718657742333\\
0.0998211616023656	-0.253898442705628\\
-0.0248167218954752	-1.84469489302107\\
-0.0459442766260001	0.788087537609972\\
0.0203113791780736	0.5815668212585\\
0.0878829551802759	0.728861864566727\\
0.0998165675293889	-0.253847928177385\\
-0.0247786173852712	-1.84404065507388\\
-0.0459449441328773	0.787065708624757\\
0.0202864388069925	0.581801683541234\\
0.0878734126087394	0.728965425982166\\
0.10021709130476	-0.247205668731719\\
-0.0247087404685947	-1.85555057720773\\
-0.0458870795866621	0.793763091783634\\
0.0204046661248414	0.579736210189876\\
0.0880726591777564	0.732635488286345\\
0.1004275041613	-0.249269111067298\\
-0.0246303405480671	-1.85641004940558\\
-0.0457597346040263	0.795074500722073\\
0.0205142349225996	0.578649269273974\\
0.0881135714084291	0.732196930176058\\
0.100441992729782	-0.249425534904074\\
-0.0246256472342556	-1.85647179386852\\
-0.0457333533321634	0.79546059576272\\
0.0205396896602432	0.578398441454283\\
0.088123116622489	0.73209466627443\\
0.100445374342665	-0.249462043490443\\
-0.024624534003121	-1.85648591907137\\
-0.0457270726484587	0.795552229918251\\
0.0205457428482101	0.578338789034525\\
0.0881253863225758	0.73207034805552\\
0.100446178352532	-0.249470725088824\\
-0.0246242680148323	-1.85648925565119\\
-0.0457249837532522	0.79558354934616\\
0.0205477763436004	0.57831875965459\\
0.0881261490081897	0.73206217578251\\
0.100446448472129	-0.249473642791519\\
-0.0246241780068668	-1.85649036566365\\
-0.0457252180363022	0.795579018761943\\
0.0205475236941188	0.578321233315645\\
0.0881260538312401	0.732063193511643\\
0.100446414654457	-0.249473279150229\\
-0.024624189889895	-1.85649023556243\\
-0.0457251409508319	0.795580368311626\\
0.0205476033580337	0.578320450012647\\
0.0881260836992177	0.732062872346924\\
0.100446425162993	-0.249473393845037\\
-0.0246241864864281	-1.85649027959671\\
-0.0457251211852506	0.795580658037551\\
0.0205476223987641	0.578320261316201\\
0.0881260907729195	0.732062795440384\\
0.10856178396298	-0.15155877093357\\
-0.00255212190401262	-1.75227612889841\\
-0.0603890687030734	0.213887694984646\\
-0.00731159168900338	0.741831114020356\\
0.0739269378322205	0.856539252757174\\
0.103391668447089	-0.0846404030239185\\
-0.0020217200804791	-1.702174285666\\
-0.0599889142220113	0.178001825552888\\
-0.00827343221899095	0.74324902668133\\
0.0733367907955527	0.861787952858752\\
0.103179359530218	-0.0818668061436522\\
-0.000860124163727591	-1.6811130025941\\
-0.0595977212101289	0.150973801489224\\
-0.00893822525482332	0.743813352937467\\
0.0729146809066775	0.865458100433001\\
0.10302462792162	-0.0798743808812867\\
-9.53246635627403e-05	-1.66710862564196\\
-0.0593394471895579	0.133097755348896\\
-0.00937821561190168	0.744189153947628\\
0.0726353881389454	0.867886936320728\\
0.102922263764363	-0.0785561209680612\\
0.00041085430862671	-1.65784035155343\\
-0.0591685119444636	0.121266848684857\\
-0.00966941337688523	0.744437860748663\\
0.0724505441733024	0.869494407771972\\
0.102854516162038	-0.0776836586312704\\
0.000745991845980412	-1.65170410846048\\
-0.059055338186774	0.113433812806454\\
-0.00986220996474418	0.744602522126286\\
0.0723281623468874	0.870558684733332\\
0.102809661681397	-0.0771060171936084\\
0.000967880245198754	-1.64764141423481\\
-0.0589804078054898	0.108247705431752\\
-0.00998985690985328	0.744711540509185\\
0.0722471356427301	0.87126332223811\\
0.102779964362089	-0.0767235717019669\\
0.0011147881491444	-1.64495158445364\\
-0.0589307990680688	0.104814066602553\\
-0.0100743708924906	0.744783728031817\\
0.0721934887830689	0.871729856076849\\
0.102442638201681	-0.082134001848154\\
0.000745393688686813	-1.64037843596966\\
-0.0479180083140558	0.292243904307435\\
0.00376160876676958	0.665055819026246\\
0.0785495827319837	0.799526819082987\\
0.104178257581282	-0.110595587501414\\
-0.00433495275164063	-1.73487502985731\\
-0.0490414934923074	0.421794380292067\\
0.00744084074043398	0.657973316296087\\
0.0807389902491103	0.779451428784629\\
0.104939778411263	-0.120896802151058\\
-0.00418335808490941	-1.73811402591463\\
-0.0490436344360176	0.421410246205778\\
0.00742631543022516	0.658025267933668\\
0.0807311654379162	0.77952819632277\\
0.104937228081377	-0.120860507002943\\
-0.0041838045255424	-1.73810336591359\\
-0.0490437688598849	0.421408276656667\\
0.00742615937594281	0.658026232371148\\
0.0807310952290799	0.779528978794576\\
0.104937208386609	-0.120860191200396\\
-0.00418380676911237	-1.73810328741132\\
-0.0490437676712131	0.421408281068344\\
0.00742616030407897	0.658026225059012\\
0.0807310955851838	0.779528974176382\\
0.104937208456926	-0.120860192858359\\
-0.00418380678637732	-1.73810328775616\\
-0.0490437677012781	0.421408281086925\\
0.00742616027818699	0.658026225116089\\
0.080731095563583	0.779528974209531\\
0.104937208437557	-0.120860192843454\\
-0.00418380680473351	-1.73810328774929\\
-0.0490437677189961	0.421408281092949\\
0.00742616026105441	0.658026225121796\\
0.0807310955470124	0.77952897421506\\
0.104937208421531	-0.120860192838087\\
-0.00418380682016947	-1.73810328774307\\
-0.0490437677339064	0.421408281097529\\
0.007426160246598	0.658026225126283\\
0.0807310955330099	0.77952897421961\\
};
\label{trajectory.x}

\addplot [color=red!70!black,densely dashed,mark=asterisk,mark options={solid},forget plot]
  table[row sep=crcr]{%
-0.683717323210604	3.68815169182611\\
-0.327632549867071	3.45999883588644\\
-0.116358749288841	1.27589524590819\\
-0.0401945425709293	0.442507074617809\\
-0.0138422445225256	0.152459248569911\\
-0.00476537104705232	0.0524887448486233\\
-0.00164047653088875	0.018069311538463\\
-0.000564731702535457	0.00622032140480202\\
-0.000194409020214836	0.00214132981944309\\
-6.69275122648013e-05	0.000737129158112723\\
-2.30423036491049e-05	0.000253766330134279\\
-7.93421397229736e-06	8.73629161413424e-05\\
-2.73257772604393e-06	3.00831797990605e-05\\
-9.31394844369653e-07	1.05197315011378e-05\\
-2.74822836255352e-07	4.10734474650512e-06\\
-5.94035213326787e-08	9.46891785082846e-07\\
-9.80485288172335e-09	2.17277390477932e-07\\
1.51499050518602e-09	4.88744443534501e-08\\
4.00171467982777e-09	1.00381093787771e-08\\
4.45336652143615e-09	1.11365108508188e-09\\
4.43928344511272e-09	-9.06387760632897e-10\\
4.32164408949004e-09	-1.33374200325783e-09\\
4.1839323323887e-09	-1.39469805116929e-09\\
4.04529364768467e-09	-1.37232744658797e-09\\
3.91002387513751e-09	-1.331941130275e-09\\
3.77899465593884e-09	-1.28857660007219e-09\\
3.65229110813259e-09	-1.24566598870161e-09\\
3.5298206573814e-09	-1.20396340111511e-09\\
3.41145346847484e-09	-1.16360594379461e-09\\
3.29705474085833e-09	-1.12458951603521e-09\\
3.18649204280474e-09	-1.08687861531423e-09\\
3.07963688679299e-09	-1.0504316480894e-09\\
2.97636497929449e-09	-1.01520673449957e-09\\
2.87655616833748e-09	-9.81163010953086e-10\\
2.78009432498897e-09	-9.48260894563246e-10\\
2.59676636009209e-09	-8.85729657184079e-10\\
2.42552760465089e-09	-8.27321921165188e-10\\
2.26558085907631e-09	-7.72765770752951e-10\\
2.12188463876841e-09	-6.28520605972302e-10\\
2.00319305354222e-09	-5.84212808876683e-10\\
1.89129539158751e-09	-5.28054015209002e-10\\
1.79300203620507e-09	-4.69448754571717e-10\\
1.65737294011937e-09	-4.31245778898148e-10\\
1.53442863980151e-09	-3.93455249519098e-10\\
1.42105638836774e-09	-3.62265308245934e-10\\
1.31648605293206e-09	-3.35480855472785e-10\\
1.21969267789507e-09	-3.10304339611537e-10\\
1.10168139426048e-09	-2.80208534209658e-10\\
9.95114670609002e-10	-2.53087765640691e-10\\
8.76291246736463e-10	-2.22854925757943e-10\\
7.71658294533759e-10	-1.96242485667248e-10\\
6.62456118687015e-10	-1.68470866961118e-10\\
5.54427236385637e-10	-1.40997772583975e-10\\
4.52363282623315e-10	-1.15041634073507e-10\\
3.50784854800701e-10	-8.92089708751058e-11\\
2.46942681471438e-10	-7.61966870012063e-11\\
1.501536258306e-10	-5.03569148442e-11\\
7.67128340431511e-11	-2.57271298900501e-11\\
};
\label{trajectory.z}

\coordinate (pt1) at (axis cs:0.03,-0.7);
\end{axis}

\begin{axis}[%
width=\picwidth\figurewidth,
height=\figureheight,
at={(\picwidthh\figurewidth,1\figureheight)},
scale only axis,
xmin=-0.2,
xmax=0.2,
xtick={-0.2,-0.1,0,0.1,0.2},
ymin=-3,
ymax=2,
ytick={-3,-2,...,2},
minor x tick num={1},
minor y tick num={1},
yticklabel style = {font=\footnotesize},
xticklabel style = {font=\footnotesize},
axis background/.style={fill=white},
xmajorgrids,
ymajorgrids,
tick label style={/pgf/number format/fixed},
]
\addplot [color=black,solid,mark=o,mark options={solid},forget plot]
table[row sep=crcr]{%
-1	15\\
-0.219739665635995	3.35455608422688\\
-0.0706565690469134	0.342465256627355\\
0.00211266487154647	0.958346046037598\\
0.0536518911113192	0.241415387649937\\
0.100010630998353	0.596327877935451\\
0.104016667947759	-0.300070142439352\\
0.0386697347797844	-0.8651575159184\\
0.0396657077396008	0.544578539695769\\
0.0665762622653818	0.0990523969273757\\
0.104459692800299	0.547322608693661\\
0.105548240936322	-0.316939966563813\\
-0.0234104773060279	-1.87550555610258\\
-0.0454529876027668	0.792231862493189\\
0.0207099735690577	0.577547410335864\\
0.0880290858580497	0.727266866762162\\
0.099867225534307	-0.254405274040538\\
-0.0250381316483922	-1.84867685952447\\
-0.0459387040830676	0.794162404364296\\
0.0204606281683812	0.580160468015114\\
0.0879400386892054	0.728242323848081\\
0.0998364403403215	-0.254066446297423\\
-0.0249423849177627	-1.84685366512932\\
-0.0459420642852919	0.791458283097746\\
0.0203936571334545	0.580792005030567\\
0.0879144356774874	0.728520216967713\\
0.0998275274763142	-0.253968440258361\\
-0.0248694087374712	-1.84559963026487\\
-0.0459433523579183	0.78950051207318\\
0.0203458671893661	0.581242047796113\\
0.0878961507589464	0.728718657742333\\
0.0998211616023656	-0.253898442705628\\
-0.0248167218954752	-1.84469489302107\\
-0.0459442766260001	0.788087537609972\\
0.0203113791780736	0.5815668212585\\
0.0878829551802759	0.728861864566727\\
0.0998165675293889	-0.253847928177385\\
-0.0247786173852712	-1.84404065507388\\
-0.0459449441328773	0.787065708624757\\
0.0202864388069925	0.581801683541234\\
0.0878734126087394	0.728965425982166\\
0.10021709130476	-0.247205668731719\\
-0.0247087404685947	-1.85555057720773\\
-0.0458870795866621	0.793763091783634\\
0.0204046661248414	0.579736210189876\\
0.0880726591777564	0.732635488286345\\
0.1004275041613	-0.249269111067298\\
-0.0246303405480671	-1.85641004940558\\
-0.0457597346040263	0.795074500722073\\
0.0205142349225996	0.578649269273974\\
0.0881135714084291	0.732196930176058\\
0.100441992729782	-0.249425534904074\\
-0.0246256472342556	-1.85647179386852\\
-0.0457333533321634	0.79546059576272\\
0.0205396896602432	0.578398441454283\\
0.088123116622489	0.73209466627443\\
0.100445374342665	-0.249462043490443\\
-0.024624534003121	-1.85648591907137\\
-0.0457270726484587	0.795552229918251\\
0.0205457428482101	0.578338789034525\\
0.0881253863225758	0.73207034805552\\
0.100446178352532	-0.249470725088824\\
-0.0246242680148323	-1.85648925565119\\
-0.0457249837532522	0.79558354934616\\
0.0205477763436004	0.57831875965459\\
0.0881261490081897	0.73206217578251\\
0.100446448472129	-0.249473642791519\\
-0.0246241780068668	-1.85649036566365\\
-0.0457252180363022	0.795579018761943\\
0.0205475236941188	0.578321233315645\\
0.0881260538312401	0.732063193511643\\
0.100446414654457	-0.249473279150229\\
-0.024624189889895	-1.85649023556243\\
-0.0457251409508319	0.795580368311626\\
0.0205476033580337	0.578320450012647\\
0.0881260836992177	0.732062872346924\\
0.100446425162993	-0.249473393845037\\
-0.0246241864864281	-1.85649027959671\\
-0.0457251211852506	0.795580658037551\\
0.0205476223987641	0.578320261316201\\
0.0881260907729195	0.732062795440384\\
0.10856178396298	-0.15155877093357\\
-0.00255212190401262	-1.75227612889841\\
-0.0603890687030734	0.213887694984646\\
-0.00731159168900338	0.741831114020356\\
0.0739269378322205	0.856539252757174\\
0.103391668447089	-0.0846404030239185\\
-0.0020217200804791	-1.702174285666\\
-0.0599889142220113	0.178001825552888\\
-0.00827343221899095	0.74324902668133\\
0.0733367907955527	0.861787952858752\\
0.103179359530218	-0.0818668061436522\\
-0.000860124163727591	-1.6811130025941\\
-0.0595977212101289	0.150973801489224\\
-0.00893822525482332	0.743813352937467\\
0.0729146809066775	0.865458100433001\\
0.10302462792162	-0.0798743808812867\\
-9.53246635627403e-05	-1.66710862564196\\
-0.0593394471895579	0.133097755348896\\
-0.00937821561190168	0.744189153947628\\
0.0726353881389454	0.867886936320728\\
0.102922263764363	-0.0785561209680612\\
0.00041085430862671	-1.65784035155343\\
-0.0591685119444636	0.121266848684857\\
-0.00966941337688523	0.744437860748663\\
0.0724505441733024	0.869494407771972\\
0.102854516162038	-0.0776836586312704\\
0.000745991845980412	-1.65170410846048\\
-0.059055338186774	0.113433812806454\\
-0.00986220996474418	0.744602522126286\\
0.0723281623468874	0.870558684733332\\
0.102809661681397	-0.0771060171936084\\
0.000967880245198754	-1.64764141423481\\
-0.0589804078054898	0.108247705431752\\
-0.00998985690985328	0.744711540509185\\
0.0722471356427301	0.87126332223811\\
0.102779964362089	-0.0767235717019669\\
0.0011147881491444	-1.64495158445364\\
-0.0589307990680688	0.104814066602553\\
-0.0100743708924906	0.744783728031817\\
0.0721934887830689	0.871729856076849\\
0.102442638201681	-0.082134001848154\\
0.000745393688686813	-1.64037843596966\\
-0.0479180083140558	0.292243904307435\\
0.00376160876676958	0.665055819026246\\
0.0785495827319837	0.799526819082987\\
0.104178257581282	-0.110595587501414\\
-0.00433495275164063	-1.73487502985731\\
-0.0490414934923074	0.421794380292067\\
0.00744084074043398	0.657973316296087\\
0.0807389902491103	0.779451428784629\\
0.104939778411263	-0.120896802151058\\
-0.00418335808490941	-1.73811402591463\\
-0.0490436344360176	0.421410246205778\\
0.00742631543022516	0.658025267933668\\
0.0807311654379162	0.77952819632277\\
0.104937228081377	-0.120860507002943\\
-0.0041838045255424	-1.73810336591359\\
-0.0490437688598849	0.421408276656667\\
0.00742615937594281	0.658026232371148\\
0.0807310952290799	0.779528978794576\\
0.104937208386609	-0.120860191200396\\
-0.00418380676911237	-1.73810328741132\\
-0.0490437676712131	0.421408281068344\\
0.00742616030407897	0.658026225059012\\
0.0807310955851838	0.779528974176382\\
0.104937208456926	-0.120860192858359\\
-0.00418380678637732	-1.73810328775616\\
-0.0490437677012781	0.421408281086925\\
0.00742616027818699	0.658026225116089\\
0.080731095563583	0.779528974209531\\
0.104937208437557	-0.120860192843454\\
-0.00418380680473351	-1.73810328774929\\
-0.0490437677189961	0.421408281092949\\
0.00742616026105441	0.658026225121796\\
0.0807310955470124	0.77952897421506\\
0.104937208421531	-0.120860192838087\\
-0.00418380682016947	-1.73810328774307\\
-0.0490437677339064	0.421408281097529\\
0.007426160246598	0.658026225126283\\
0.0807310955330099	0.77952897421961\\
};

\addplot [color=red!70!black,densely dashed,mark=asterisk,mark options={solid},forget plot]
table[row sep=crcr]{%
-0.683717323210604	3.68815169182611\\
-0.327632549867071	3.45999883588644\\
-0.116358749288841	1.27589524590819\\
-0.0401945425709293	0.442507074617809\\
-0.0138422445225256	0.152459248569911\\
-0.00476537104705232	0.0524887448486233\\
-0.00164047653088875	0.018069311538463\\
-0.000564731702535457	0.00622032140480202\\
-0.000194409020214836	0.00214132981944309\\
-6.69275122648013e-05	0.000737129158112723\\
-2.30423036491049e-05	0.000253766330134279\\
-7.93421397229736e-06	8.73629161413424e-05\\
-2.73257772604393e-06	3.00831797990605e-05\\
-9.31394844369653e-07	1.05197315011378e-05\\
-2.74822836255352e-07	4.10734474650512e-06\\
-5.94035213326787e-08	9.46891785082846e-07\\
-9.80485288172335e-09	2.17277390477932e-07\\
1.51499050518602e-09	4.88744443534501e-08\\
4.00171467982777e-09	1.00381093787771e-08\\
4.45336652143615e-09	1.11365108508188e-09\\
4.43928344511272e-09	-9.06387760632897e-10\\
4.32164408949004e-09	-1.33374200325783e-09\\
4.1839323323887e-09	-1.39469805116929e-09\\
4.04529364768467e-09	-1.37232744658797e-09\\
3.91002387513751e-09	-1.331941130275e-09\\
3.77899465593884e-09	-1.28857660007219e-09\\
3.65229110813259e-09	-1.24566598870161e-09\\
3.5298206573814e-09	-1.20396340111511e-09\\
3.41145346847484e-09	-1.16360594379461e-09\\
3.29705474085833e-09	-1.12458951603521e-09\\
3.18649204280474e-09	-1.08687861531423e-09\\
3.07963688679299e-09	-1.0504316480894e-09\\
2.97636497929449e-09	-1.01520673449957e-09\\
2.87655616833748e-09	-9.81163010953086e-10\\
2.78009432498897e-09	-9.48260894563246e-10\\
2.59676636009209e-09	-8.85729657184079e-10\\
2.42552760465089e-09	-8.27321921165188e-10\\
2.26558085907631e-09	-7.72765770752951e-10\\
2.12188463876841e-09	-6.28520605972302e-10\\
2.00319305354222e-09	-5.84212808876683e-10\\
1.89129539158751e-09	-5.28054015209002e-10\\
1.79300203620507e-09	-4.69448754571717e-10\\
1.65737294011937e-09	-4.31245778898148e-10\\
1.53442863980151e-09	-3.93455249519098e-10\\
1.42105638836774e-09	-3.62265308245934e-10\\
1.31648605293206e-09	-3.35480855472785e-10\\
1.21969267789507e-09	-3.10304339611537e-10\\
1.10168139426048e-09	-2.80208534209658e-10\\
9.95114670609002e-10	-2.53087765640691e-10\\
8.76291246736463e-10	-2.22854925757943e-10\\
7.71658294533759e-10	-1.96242485667248e-10\\
6.62456118687015e-10	-1.68470866961118e-10\\
5.54427236385637e-10	-1.40997772583975e-10\\
4.52363282623315e-10	-1.15041634073507e-10\\
3.50784854800701e-10	-8.92089708751058e-11\\
2.46942681471438e-10	-7.61966870012063e-11\\
1.501536258306e-10	-5.03569148442e-11\\
7.67128340431511e-11	-2.57271298900501e-11\\
};
\coordinate (pt2) at (axis description cs:0,0.6);
\end{axis}

\node[square node, name=z2, minimum height=0.7cm,minimum width=0.7cm] at (pt1) {};
\draw [-latex] (z2.north east) to (pt2);
\end{tikzpicture}%
\vspace*{-0.4cm}
\caption{State trajectory from $x_{0}=[-1\:\:15]^{\top}$, \ref{trajectory.x} $x(i)$, \ref{trajectory.z} $z(i)$.} 
\label{fig.states}
\end{center}
\end{figure}

Figure~\ref{fig.states} shows the closed-loop state trajectories for the true plant and the undisturbed nominal model generated by the inputs in Figure~\ref{fig.inputs}. As expected, the undisturbed state converges to the origin fairly fast, but the true state remains disturbed by $\hat{w}(i)$.

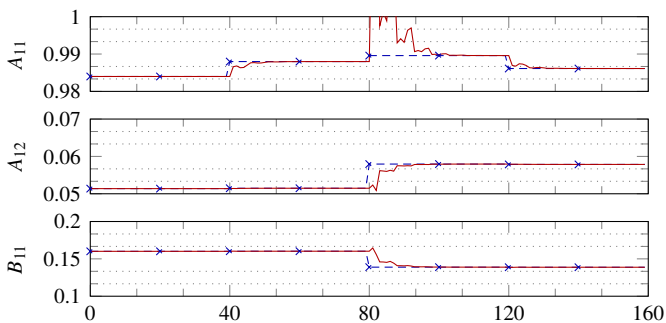
\begin{figure}
\begin{center}
\setlength\figureheight{0.2\textwidth}
\setlength\figurewidth{0.4\textwidth}
\pgfplotsset{major grid style={dashed,gray}}
\pgfplotsset{minor grid style={dotted,gray}}
\newcommand{\picheight}{0.27}

\begin{tikzpicture}[align=center,
square node/.style={rectangle,draw=black,fill=none,thin,minimum width=2.5mm,minimum height=2.5mm},
every pin/.style={fill=white}]
\begin{axis}[%
width=\figurewidth,
height=\picheight\figureheight,
at={(0\figurewidth,1\figureheight)},
scale only axis,
xmin=0,
xmax=160,
xtick={0,40,...,160},
xticklabels={,,},
ymin=0.98,
ymax=1,
ytick={0.98,0.99,1},
minor x tick num={2},
minor y tick num={2},
xticklabel style = {font=\footnotesize},
yticklabel style = {font=\footnotesize},
yminorgrids,
ylabel={\footnotesize$A_{11}$},
tick label style={/pgf/number format/fixed},
]
\addplot [color=blue!70!black,densely dashed,mark=x,mark repeat={20},forget plot]
  table[row sep=crcr]{%
0	0.983990057302899\\
1	0.983990057302899\\
2	0.983990057302899\\
3	0.983990057302899\\
4	0.983990057302899\\
5	0.983990057302899\\
6	0.983990057302899\\
7	0.983990057302899\\
8	0.983990057302899\\
9	0.983990057302899\\
10	0.983990057302899\\
11	0.983990057302899\\
12	0.983990057302899\\
13	0.983990057302899\\
14	0.983990057302899\\
15	0.983990057302899\\
16	0.983990057302899\\
17	0.983990057302899\\
18	0.983990057302899\\
19	0.983990057302899\\
20	0.983990057302899\\
21	0.983990057302899\\
22	0.983990057302899\\
23	0.983990057302899\\
24	0.983990057302899\\
25	0.983990057302899\\
26	0.983990057302899\\
27	0.983990057302899\\
28	0.983990057302899\\
29	0.983990057302899\\
30	0.983990057302899\\
31	0.983990057302899\\
32	0.983990057302899\\
33	0.983990057302899\\
34	0.983990057302899\\
35	0.983990057302899\\
36	0.983990057302899\\
37	0.983990057302899\\
38	0.983990057302899\\
39	0.983990057302899\\
40	0.987981378156579\\
41	0.987981378156579\\
42	0.987981378156579\\
43	0.987981378156579\\
44	0.987981378156579\\
45	0.987981378156579\\
46	0.987981378156579\\
47	0.987981378156579\\
48	0.987981378156579\\
49	0.987981378156579\\
50	0.987981378156579\\
51	0.987981378156579\\
52	0.987981378156579\\
53	0.987981378156579\\
54	0.987981378156579\\
55	0.987981378156579\\
56	0.987981378156579\\
57	0.987981378156579\\
58	0.987981378156579\\
59	0.987981378156579\\
60	0.987981378156579\\
61	0.987981378156579\\
62	0.987981378156579\\
63	0.987981378156579\\
64	0.987981378156579\\
65	0.987981378156579\\
66	0.987981378156579\\
67	0.987981378156579\\
68	0.987981378156579\\
69	0.987981378156579\\
70	0.987981378156579\\
71	0.987981378156579\\
72	0.987981378156579\\
73	0.987981378156579\\
74	0.987981378156579\\
75	0.987981378156579\\
76	0.987981378156579\\
77	0.987981378156579\\
78	0.987981378156579\\
79	0.987981378156579\\
80	0.989582295733778\\
81	0.989582295733778\\
82	0.989582295733778\\
83	0.989582295733778\\
84	0.989582295733778\\
85	0.989582295733778\\
86	0.989582295733778\\
87	0.989582295733778\\
88	0.989582295733778\\
89	0.989582295733778\\
90	0.989582295733778\\
91	0.989582295733778\\
92	0.989582295733778\\
93	0.989582295733778\\
94	0.989582295733778\\
95	0.989582295733778\\
96	0.989582295733778\\
97	0.989582295733778\\
98	0.989582295733778\\
99	0.989582295733778\\
100	0.989582295733778\\
101	0.989582295733778\\
102	0.989582295733778\\
103	0.989582295733778\\
104	0.989582295733778\\
105	0.989582295733778\\
106	0.989582295733778\\
107	0.989582295733778\\
108	0.989582295733778\\
109	0.989582295733778\\
110	0.989582295733778\\
111	0.989582295733778\\
112	0.989582295733778\\
113	0.989582295733778\\
114	0.989582295733778\\
115	0.989582295733778\\
116	0.989582295733778\\
117	0.989582295733778\\
118	0.989582295733778\\
119	0.989582295733778\\
120	0.986120310444952\\
121	0.986120310444952\\
122	0.986120310444952\\
123	0.986120310444952\\
124	0.986120310444952\\
125	0.986120310444952\\
126	0.986120310444952\\
127	0.986120310444952\\
128	0.986120310444952\\
129	0.986120310444952\\
130	0.986120310444952\\
131	0.986120310444952\\
132	0.986120310444952\\
133	0.986120310444952\\
134	0.986120310444952\\
135	0.986120310444952\\
136	0.986120310444952\\
137	0.986120310444952\\
138	0.986120310444952\\
139	0.986120310444952\\
140	0.986120310444952\\
141	0.986120310444952\\
142	0.986120310444952\\
143	0.986120310444952\\
144	0.986120310444952\\
145	0.986120310444952\\
146	0.986120310444952\\
147	0.986120310444952\\
148	0.986120310444952\\
149	0.986120310444952\\
150	0.986120310444952\\
151	0.986120310444952\\
152	0.986120310444952\\
153	0.986120310444952\\
154	0.986120310444952\\
155	0.986120310444952\\
156	0.986120310444952\\
157	0.986120310444952\\
158	0.986120310444952\\
159	0.986120310444952\\
};\label{param.true}
\addplot [color=red!70!black,solid,forget plot]
  table[row sep=crcr]{%
0	0.983990057302899\\
1	0.983990057302899\\
2	0.983990057302899\\
3	0.983990057302898\\
4	0.983990057302899\\
5	0.983990057302899\\
6	0.983990057302899\\
7	0.983990057302899\\
8	0.983990057302899\\
9	0.983990057302899\\
10	0.983990057302899\\
11	0.983990057302899\\
12	0.983990057302899\\
13	0.983990057302899\\
14	0.983990057302899\\
15	0.983990057302899\\
16	0.983990057302899\\
17	0.983990057302899\\
18	0.983990057302899\\
19	0.983990057302899\\
20	0.983990057302899\\
21	0.983990057302899\\
22	0.983990057302899\\
23	0.983990057302899\\
24	0.983990057302899\\
25	0.983990057302899\\
26	0.983990057302899\\
27	0.983990057302899\\
28	0.983990057302899\\
29	0.983990057302899\\
30	0.983990057302899\\
31	0.983990057302899\\
32	0.983990057302899\\
33	0.983990057302899\\
34	0.983990057302899\\
35	0.983990057302899\\
36	0.983990057302899\\
37	0.983990057302899\\
38	0.983990057302899\\
39	0.983990057302899\\
40	0.983990057302899\\
41	0.986609090038827\\
42	0.986804647455349\\
43	0.986332622094454\\
44	0.986400441455745\\
45	0.987030901377162\\
46	0.9876541130815\\
47	0.987701207106375\\
48	0.987590862545638\\
49	0.987606568851457\\
50	0.987756218363829\\
51	0.987903696753097\\
52	0.9879148615415\\
53	0.987888834104705\\
54	0.987892530110393\\
55	0.987927992126234\\
56	0.987962944772699\\
57	0.987965593752401\\
58	0.987959428970109\\
59	0.98796030366899\\
60	0.987968713254549\\
61	0.987977003976821\\
62	0.987977632574339\\
63	0.987976170451158\\
64	0.987976377837764\\
65	0.987978373012166\\
66	0.987980340158832\\
67	0.98798048932929\\
68	0.987980142419448\\
69	0.987980191624183\\
70	0.987980665042432\\
71	0.9879811318358\\
72	0.987981167234236\\
73	0.987981084913926\\
74	0.987981096589838\\
75	0.987981208932324\\
76	0.987981319703561\\
77	0.987981328103775\\
78	0.987981308568972\\
79	0.98798131133968\\
80	0.987981337998944\\
81	1.04047962431082\\
82	1.05595922085732\\
83	0.997511027306028\\
84	1.00116049218471\\
85	0.998875003070065\\
86	1.01184042810141\\
87	1.01422715595961\\
88	0.99298668436743\\
89	0.994102184767754\\
90	0.993132515216929\\
91	0.996357134827036\\
92	0.996987386062378\\
93	0.990647144580672\\
94	0.990941317261654\\
95	0.990630520025666\\
96	0.991389603634853\\
97	0.991542970085076\\
98	0.98987381904639\\
99	0.989947286795263\\
100	0.989861361870424\\
101	0.990039426799555\\
102	0.990075977057503\\
103	0.989657451792109\\
104	0.989675389515795\\
105	0.989653153055496\\
106	0.989695040788818\\
107	0.989703710045225\\
108	0.989601070181839\\
109	0.989605400920141\\
110	0.989599836435795\\
111	0.989609715952371\\
112	0.98961177073738\\
113	0.989586899882656\\
114	0.989587938844206\\
115	0.989586572959859\\
116	0.989588907587515\\
117	0.989589394698138\\
118	0.989583411927758\\
119	0.989583660217121\\
120	0.989583328897341\\
121	0.986935671984692\\
122	0.98675670590945\\
123	0.987392850219336\\
124	0.987300549949579\\
125	0.986927848538365\\
126	0.986325176805786\\
127	0.986269076318036\\
128	0.986405336788656\\
129	0.986383567708325\\
130	0.986302717310315\\
131	0.986172854632364\\
132	0.986156915956949\\
133	0.986183806773789\\
134	0.98617923940306\\
135	0.986162108192146\\
136	0.986132838388386\\
137	0.986129067481509\\
138	0.986135169421272\\
139	0.986134114911911\\
140	0.986130150923915\\
141	0.98612328615969\\
142	0.986122392066741\\
143	0.98612382503966\\
144	0.986123576405656\\
145	0.986122641284911\\
146	0.986121016744757\\
147	0.986120804617512\\
148	0.986121143829892\\
149	0.986121084917709\\
150	0.986120863319883\\
151	0.986120478061483\\
152	0.986120427725272\\
153	0.986120508174835\\
154	0.986120494199748\\
155	0.986120441631017\\
156	0.986120350221605\\
157	0.986120338276732\\
158	0.986120357365139\\
159	0.986120354049069\\
}; \label{param.estim}
\end{axis}

\begin{axis}[%
width=\figurewidth,
height=\picheight\figureheight,
at={(0\figurewidth,0.63\figureheight)},
scale only axis,
tick label style={/pgf/number format/fixed},
xmin=0,
xmax=160,
xtick={0,40,...,160},
xticklabels={,,},
ymin=0.05,
ymax=0.07,
ytick={0.05,0.06,0.07},
minor x tick num={2},
minor y tick num={2},
xticklabel style = {font=\footnotesize},
yticklabel style = {font=\footnotesize},
yminorgrids,
ylabel={\footnotesize$A_{12}$},
tick label style={/pgf/number format/fixed},
scaled y ticks=false,
]
\addplot [color=blue!70!black,densely dashed,mark=x,mark repeat={20},forget plot]
  table[row sep=crcr]{%
0	0.0513761152486467\\
1	0.0513761152486467\\
2	0.0513761152486467\\
3	0.0513761152486467\\
4	0.0513761152486467\\
5	0.0513761152486467\\
6	0.0513761152486467\\
7	0.0513761152486467\\
8	0.0513761152486467\\
9	0.0513761152486467\\
10	0.0513761152486467\\
11	0.0513761152486467\\
12	0.0513761152486467\\
13	0.0513761152486467\\
14	0.0513761152486467\\
15	0.0513761152486467\\
16	0.0513761152486467\\
17	0.0513761152486467\\
18	0.0513761152486467\\
19	0.0513761152486467\\
20	0.0513761152486467\\
21	0.0513761152486467\\
22	0.0513761152486467\\
23	0.0513761152486467\\
24	0.0513761152486467\\
25	0.0513761152486467\\
26	0.0513761152486467\\
27	0.0513761152486467\\
28	0.0513761152486467\\
29	0.0513761152486467\\
30	0.0513761152486467\\
31	0.0513761152486467\\
32	0.0513761152486467\\
33	0.0513761152486467\\
34	0.0513761152486467\\
35	0.0513761152486467\\
36	0.0513761152486467\\
37	0.0513761152486467\\
38	0.0513761152486467\\
39	0.0513761152486467\\
40	0.0514797254482036\\
41	0.0514797254482036\\
42	0.0514797254482036\\
43	0.0514797254482036\\
44	0.0514797254482036\\
45	0.0514797254482036\\
46	0.0514797254482036\\
47	0.0514797254482036\\
48	0.0514797254482036\\
49	0.0514797254482036\\
50	0.0514797254482036\\
51	0.0514797254482036\\
52	0.0514797254482036\\
53	0.0514797254482036\\
54	0.0514797254482036\\
55	0.0514797254482036\\
56	0.0514797254482036\\
57	0.0514797254482036\\
58	0.0514797254482036\\
59	0.0514797254482036\\
60	0.0514797254482036\\
61	0.0514797254482036\\
62	0.0514797254482036\\
63	0.0514797254482036\\
64	0.0514797254482036\\
65	0.0514797254482036\\
66	0.0514797254482036\\
67	0.0514797254482036\\
68	0.0514797254482036\\
69	0.0514797254482036\\
70	0.0514797254482036\\
71	0.0514797254482036\\
72	0.0514797254482036\\
73	0.0514797254482036\\
74	0.0514797254482036\\
75	0.0514797254482036\\
76	0.0514797254482036\\
77	0.0514797254482036\\
78	0.0514797254482036\\
79	0.0514797254482036\\
80	0.0579498253851449\\
81	0.0579498253851449\\
82	0.0579498253851449\\
83	0.0579498253851449\\
84	0.0579498253851449\\
85	0.0579498253851449\\
86	0.0579498253851449\\
87	0.0579498253851449\\
88	0.0579498253851449\\
89	0.0579498253851449\\
90	0.0579498253851449\\
91	0.0579498253851449\\
92	0.0579498253851449\\
93	0.0579498253851449\\
94	0.0579498253851449\\
95	0.0579498253851449\\
96	0.0579498253851449\\
97	0.0579498253851449\\
98	0.0579498253851449\\
99	0.0579498253851449\\
100	0.0579498253851449\\
101	0.0579498253851449\\
102	0.0579498253851449\\
103	0.0579498253851449\\
104	0.0579498253851449\\
105	0.0579498253851449\\
106	0.0579498253851449\\
107	0.0579498253851449\\
108	0.0579498253851449\\
109	0.0579498253851449\\
110	0.0579498253851449\\
111	0.0579498253851449\\
112	0.0579498253851449\\
113	0.0579498253851449\\
114	0.0579498253851449\\
115	0.0579498253851449\\
116	0.0579498253851449\\
117	0.0579498253851449\\
118	0.0579498253851449\\
119	0.0579498253851449\\
120	0.0578553493333264\\
121	0.0578553493333264\\
122	0.0578553493333264\\
123	0.0578553493333264\\
124	0.0578553493333264\\
125	0.0578553493333264\\
126	0.0578553493333264\\
127	0.0578553493333264\\
128	0.0578553493333264\\
129	0.0578553493333264\\
130	0.0578553493333264\\
131	0.0578553493333264\\
132	0.0578553493333264\\
133	0.0578553493333264\\
134	0.0578553493333264\\
135	0.0578553493333264\\
136	0.0578553493333264\\
137	0.0578553493333264\\
138	0.0578553493333264\\
139	0.0578553493333264\\
140	0.0578553493333264\\
141	0.0578553493333264\\
142	0.0578553493333264\\
143	0.0578553493333264\\
144	0.0578553493333264\\
145	0.0578553493333264\\
146	0.0578553493333264\\
147	0.0578553493333264\\
148	0.0578553493333264\\
149	0.0578553493333264\\
150	0.0578553493333264\\
151	0.0578553493333264\\
152	0.0578553493333264\\
153	0.0578553493333264\\
154	0.0578553493333264\\
155	0.0578553493333264\\
156	0.0578553493333264\\
157	0.0578553493333264\\
158	0.0578553493333264\\
159	0.0578553493333264\\
};
\addplot [color=red!70!black,solid,forget plot]
  table[row sep=crcr]{%
0	0.0513761152486467\\
1	0.0513761152486467\\
2	0.0513761152486467\\
3	0.0513761152486466\\
4	0.0513761152486467\\
5	0.0513761152486467\\
6	0.0513761152486467\\
7	0.0513761152486467\\
8	0.0513761152486467\\
9	0.0513761152486467\\
10	0.0513761152486467\\
11	0.0513761152486467\\
12	0.0513761152486467\\
13	0.0513761152486467\\
14	0.0513761152486467\\
15	0.0513761152486467\\
16	0.0513761152486467\\
17	0.0513761152486467\\
18	0.0513761152486467\\
19	0.0513761152486467\\
20	0.0513761152486467\\
21	0.0513761152486467\\
22	0.0513761152486467\\
23	0.0513761152486467\\
24	0.0513761152486467\\
25	0.0513761152486467\\
26	0.0513761152486467\\
27	0.0513761152486467\\
28	0.0513761152486467\\
29	0.0513761152486467\\
30	0.0513761152486467\\
31	0.0513761152486467\\
32	0.0513761152486467\\
33	0.0513761152486467\\
34	0.0513761152486467\\
35	0.0513761152486467\\
36	0.0513761152486467\\
37	0.0513761152486467\\
38	0.0513761152486467\\
39	0.0513761152486467\\
40	0.0513761152486467\\
41	0.0514183030652666\\
42	0.0513964243980826\\
43	0.0514414381832842\\
44	0.0514373318553235\\
45	0.0514552428800435\\
46	0.0514653226157454\\
47	0.0514601114568103\\
48	0.0514706323179933\\
49	0.0514696731859779\\
50	0.0514739264395078\\
51	0.0514763152361906\\
52	0.0514750792803942\\
53	0.0514775673423701\\
54	0.0514773408655831\\
55	0.0514783498321072\\
56	0.051478916625138\\
57	0.0514786233200733\\
58	0.0514792133085903\\
59	0.0514791596474052\\
60	0.0514793990321757\\
61	0.0514795335357032\\
62	0.0514794639300051\\
63	0.0514796039139771\\
64	0.0514795911861536\\
65	0.0514796479893997\\
66	0.0514796799079742\\
67	0.0514796633897519\\
68	0.0514796966077187\\
69	0.0514796935876446\\
70	0.0514797070668646\\
71	0.0514797146413593\\
72	0.0514797107215118\\
73	0.051479718604221\\
74	0.0514797178875634\\
75	0.0514797210862285\\
76	0.0514797228836934\\
77	0.0514797219534939\\
78	0.0514797238240945\\
79	0.0514797236540296\\
80	0.0514797244130866\\
81	0.0523316071830172\\
82	0.0508897259694838\\
83	0.0561856631879885\\
84	0.0560921435364035\\
85	0.0559841063862846\\
86	0.0562887383197259\\
87	0.0560532118911389\\
88	0.0575009045334286\\
89	0.0574828566891782\\
90	0.0574452861598216\\
91	0.0575084608109331\\
92	0.0574552088028156\\
93	0.0578399960241058\\
94	0.0578359722293253\\
95	0.0578247468612629\\
96	0.0578386975902901\\
97	0.0578263330052214\\
98	0.0579232941204237\\
99	0.0579223506747095\\
100	0.0579193320877764\\
101	0.0579225253781187\\
102	0.0579196282401062\\
103	0.0579434566973182\\
104	0.0579432329473787\\
105	0.0579424622516982\\
106	0.0579432048666197\\
107	0.0579425230536443\\
108	0.0579483025393414\\
109	0.0579482493943395\\
110	0.0579480580326113\\
111	0.0579482320596546\\
112	0.0579480711666889\\
113	0.0579494621751927\\
114	0.0579494495555024\\
115	0.0579494028132672\\
116	0.057949443773372\\
117	0.0579494057370793\\
118	0.0579497389033192\\
119	0.0579497359076717\\
120	0.0579497246058212\\
121	0.057903393009636\\
122	0.0579171983752277\\
123	0.057885895676995\\
124	0.0578893797687783\\
125	0.0578773572521827\\
126	0.0578661620087648\\
127	0.0578709290986018\\
128	0.0578621987833032\\
129	0.0578631680302899\\
130	0.0578602260902102\\
131	0.0578576184340468\\
132	0.0578588795076146\\
133	0.0578569323971329\\
134	0.0578571544801029\\
135	0.0578564950112773\\
136	0.0578558776369359\\
137	0.0578561774434659\\
138	0.0578557227725792\\
139	0.05785577506042\\
140	0.0578556205349875\\
141	0.0578554741526885\\
142	0.0578555453239195\\
143	0.0578554378283473\\
144	0.0578554502135067\\
145	0.0578554136526831\\
146	0.0578553789229175\\
147	0.0578553958135288\\
148	0.0578553703266554\\
149	0.057855373264421\\
150	0.0578553645944656\\
151	0.0578553563533545\\
152	0.05785536036165\\
153	0.0578553543147504\\
154	0.0578553550118233\\
155	0.0578553529547454\\
156	0.0578553509991151\\
157	0.0578553519503066\\
158	0.0578553505154196\\
159	0.0578553506808342\\
};
\end{axis}

\begin{axis}[%
width=\figurewidth,
height=\picheight\figureheight,
at={(0\figurewidth,0.26\figureheight)},
scale only axis,
tick label style={/pgf/number format/fixed},
xmin=0,
xmax=160,
xtick={0,40,...,160},
ymin=0.1,
ymax=0.2,
ytick={0.1,0.15,0.2},
minor x tick num={2},
minor y tick num={2},
xticklabel style = {font=\footnotesize},
yticklabel style = {font=\footnotesize},
yminorgrids,
tick label style={/pgf/number format/fixed},
ylabel={\footnotesize$B_{11}$},
scaled y ticks=false,
]
\addplot [color=blue!70!black,densely dashed,mark=x,mark repeat={20},forget plot]
  table[row sep=crcr]{%
0	0.16009942697101\\
1	0.16009942697101\\
2	0.16009942697101\\
3	0.16009942697101\\
4	0.16009942697101\\
5	0.16009942697101\\
6	0.16009942697101\\
7	0.16009942697101\\
8	0.16009942697101\\
9	0.16009942697101\\
10	0.16009942697101\\
11	0.16009942697101\\
12	0.16009942697101\\
13	0.16009942697101\\
14	0.16009942697101\\
15	0.16009942697101\\
16	0.16009942697101\\
17	0.16009942697101\\
18	0.16009942697101\\
19	0.16009942697101\\
20	0.16009942697101\\
21	0.16009942697101\\
22	0.16009942697101\\
23	0.16009942697101\\
24	0.16009942697101\\
25	0.16009942697101\\
26	0.16009942697101\\
27	0.16009942697101\\
28	0.16009942697101\\
29	0.16009942697101\\
30	0.16009942697101\\
31	0.16009942697101\\
32	0.16009942697101\\
33	0.16009942697101\\
34	0.16009942697101\\
35	0.16009942697101\\
36	0.16009942697101\\
37	0.16009942697101\\
38	0.16009942697101\\
39	0.16009942697101\\
40	0.160248291245606\\
41	0.160248291245606\\
42	0.160248291245606\\
43	0.160248291245606\\
44	0.160248291245606\\
45	0.160248291245606\\
46	0.160248291245606\\
47	0.160248291245606\\
48	0.160248291245606\\
49	0.160248291245606\\
50	0.160248291245606\\
51	0.160248291245606\\
52	0.160248291245606\\
53	0.160248291245606\\
54	0.160248291245606\\
55	0.160248291245606\\
56	0.160248291245606\\
57	0.160248291245606\\
58	0.160248291245606\\
59	0.160248291245606\\
60	0.160248291245606\\
61	0.160248291245606\\
62	0.160248291245606\\
63	0.160248291245606\\
64	0.160248291245606\\
65	0.160248291245606\\
66	0.160248291245606\\
67	0.160248291245606\\
68	0.160248291245606\\
69	0.160248291245606\\
70	0.160248291245606\\
71	0.160248291245606\\
72	0.160248291245606\\
73	0.160248291245606\\
74	0.160248291245606\\
75	0.160248291245606\\
76	0.160248291245606\\
77	0.160248291245606\\
78	0.160248291245606\\
79	0.160248291245606\\
80	0.138902723549625\\
81	0.138902723549625\\
82	0.138902723549625\\
83	0.138902723549625\\
84	0.138902723549625\\
85	0.138902723549625\\
86	0.138902723549625\\
87	0.138902723549625\\
88	0.138902723549625\\
89	0.138902723549625\\
90	0.138902723549625\\
91	0.138902723549625\\
92	0.138902723549625\\
93	0.138902723549625\\
94	0.138902723549625\\
95	0.138902723549625\\
96	0.138902723549625\\
97	0.138902723549625\\
98	0.138902723549625\\
99	0.138902723549625\\
100	0.138902723549625\\
101	0.138902723549625\\
102	0.138902723549625\\
103	0.138902723549625\\
104	0.138902723549625\\
105	0.138902723549625\\
106	0.138902723549625\\
107	0.138902723549625\\
108	0.138902723549625\\
109	0.138902723549625\\
110	0.138902723549625\\
111	0.138902723549625\\
112	0.138902723549625\\
113	0.138902723549625\\
114	0.138902723549625\\
115	0.138902723549625\\
116	0.138902723549625\\
117	0.138902723549625\\
118	0.138902723549625\\
119	0.138902723549625\\
120	0.138796895550476\\
121	0.138796895550476\\
122	0.138796895550476\\
123	0.138796895550476\\
124	0.138796895550476\\
125	0.138796895550476\\
126	0.138796895550476\\
127	0.138796895550476\\
128	0.138796895550476\\
129	0.138796895550476\\
130	0.138796895550476\\
131	0.138796895550476\\
132	0.138796895550476\\
133	0.138796895550476\\
134	0.138796895550476\\
135	0.138796895550476\\
136	0.138796895550476\\
137	0.138796895550476\\
138	0.138796895550476\\
139	0.138796895550476\\
140	0.138796895550476\\
141	0.138796895550476\\
142	0.138796895550476\\
143	0.138796895550476\\
144	0.138796895550476\\
145	0.138796895550476\\
146	0.138796895550476\\
147	0.138796895550476\\
148	0.138796895550476\\
149	0.138796895550476\\
150	0.138796895550476\\
151	0.138796895550476\\
152	0.138796895550476\\
153	0.138796895550476\\
154	0.138796895550476\\
155	0.138796895550476\\
156	0.138796895550476\\
157	0.138796895550476\\
158	0.138796895550476\\
159	0.138796895550476\\
};
\addplot [color=red!70!black,solid,forget plot]
  table[row sep=crcr]{%
0	0.16009942697101\\
1	0.16009942697101\\
2	0.16009942697101\\
3	0.16009942697101\\
4	0.16009942697101\\
5	0.16009942697101\\
6	0.16009942697101\\
7	0.16009942697101\\
8	0.16009942697101\\
9	0.16009942697101\\
10	0.16009942697101\\
11	0.16009942697101\\
12	0.16009942697101\\
13	0.16009942697101\\
14	0.16009942697101\\
15	0.16009942697101\\
16	0.16009942697101\\
17	0.16009942697101\\
18	0.16009942697101\\
19	0.16009942697101\\
20	0.16009942697101\\
21	0.16009942697101\\
22	0.16009942697101\\
23	0.16009942697101\\
24	0.16009942697101\\
25	0.16009942697101\\
26	0.16009942697101\\
27	0.16009942697101\\
28	0.16009942697101\\
29	0.16009942697101\\
30	0.16009942697101\\
31	0.16009942697101\\
32	0.16009942697101\\
33	0.16009942697101\\
34	0.16009942697101\\
35	0.16009942697101\\
36	0.16009942697101\\
37	0.16009942697101\\
38	0.16009942697101\\
39	0.16009942697101\\
40	0.16009942697101\\
41	0.160316848360471\\
42	0.160193282443876\\
43	0.160098615236655\\
44	0.160097850732336\\
45	0.160212013340475\\
46	0.160263715940163\\
47	0.16023490156551\\
48	0.160212783942452\\
49	0.160212592053041\\
50	0.160239667232698\\
51	0.160251905722309\\
52	0.160245096845186\\
53	0.160239876877284\\
54	0.160239830162622\\
55	0.160246246313283\\
56	0.160249146426763\\
57	0.160247532437537\\
58	0.160246295744484\\
59	0.16024628455529\\
60	0.160247806198346\\
61	0.160248494041851\\
62	0.160248111144771\\
63	0.160247817808814\\
64	0.160247815145296\\
65	0.160248176163387\\
66	0.160248339362191\\
67	0.160248248505942\\
68	0.160248178905757\\
69	0.160248178273149\\
70	0.160248263937461\\
71	0.160248302663451\\
72	0.16024828110319\\
73	0.160248264587242\\
74	0.160248264437092\\
75	0.160248284765336\\
76	0.160248293955089\\
77	0.160248288838761\\
78	0.160248284919481\\
79	0.160248284883847\\
80	0.160248289707814\\
81	0.164603625741876\\
82	0.155925284165006\\
83	0.145925067010457\\
84	0.145819332236941\\
85	0.145305309171334\\
86	0.14662661046862\\
87	0.144297753914393\\
88	0.140831048994236\\
89	0.140839271438621\\
90	0.140638876537766\\
91	0.140980330147997\\
92	0.140432834245103\\
93	0.139410162969053\\
94	0.139413939244692\\
95	0.139350865769553\\
96	0.139432451640119\\
97	0.139300174848288\\
98	0.139031223236531\\
99	0.139032090707071\\
100	0.139014696145456\\
101	0.139033970829329\\
102	0.139002007136741\\
103	0.138934507209148\\
104	0.138934682339238\\
105	0.138930177694235\\
106	0.13893473006095\\
107	0.138927042419156\\
108	0.138910471933083\\
109	0.138910506526358\\
110	0.138909378139059\\
111	0.138910454524823\\
112	0.138908613401902\\
113	0.13890459510365\\
114	0.138904602036378\\
115	0.138904324837202\\
116	0.138904579608451\\
117	0.138904140023299\\
118	0.138903172893291\\
119	0.138903174321478\\
120	0.138903107044681\\
121	0.138613869107444\\
122	0.138770303931567\\
123	0.138875118278316\\
124	0.138879132037091\\
125	0.13881018730039\\
126	0.138752672684928\\
127	0.138791103449865\\
128	0.138813666686152\\
129	0.138814317981839\\
130	0.138799557930943\\
131	0.138787821259427\\
132	0.138795909394517\\
133	0.138800404317744\\
134	0.138800598412697\\
135	0.13879742310301\\
136	0.138794800190917\\
137	0.138796682321862\\
138	0.138797704848618\\
139	0.138797752722253\\
140	0.138797015435927\\
141	0.138796401442448\\
142	0.138796846050015\\
143	0.138797086320095\\
144	0.13879709777852\\
145	0.138796923707272\\
146	0.138796778470847\\
147	0.138796883864846\\
148	0.138796940749321\\
149	0.138796943473909\\
150	0.138796902215819\\
151	0.138796867776736\\
152	0.138796892780849\\
153	0.138796906272343\\
154	0.138796906919207\\
155	0.138796897131271\\
156	0.138796888960189\\
157	0.138796894893424\\
158	0.138796898094599\\
159	0.13879689824812\\
};
\end{axis}

\end{tikzpicture}%
\caption{Parameter estimation results, \ref{param.true} true plant values $\left(A,B\right)$, \ref{param.estim} estimates $\left(\mathcal{A}(i),\mathcal{B}(i)\right)$.} 
\label{fig.ID}
\end{center}
\end{figure}

A selection of the estimates is shown in Figure~\ref{fig.ID}. Given the deterministic framework, convergence to the true parameters is achieved in finite time, however the new estimates are not necessarily a feasible replacement for the prediction model. In this particular example, the model associated to a 25\% increase in the load, say $\bigl(\tilde{A},\tilde{B}\bigr)$, results in $\bigl(\tilde{A}+\tilde{B}K\bigr)\mathbb{Z}_{f}\nsubseteq\mathbb{Z}_{f}$ thus breaking the invariance of the terminal constraint set. It is easy to show, however, that if $\mathbb{Z}_{f}$ is computed as a $\lambda$-contractive set, with $\lambda=0.99$, then the invariance property holds.

\section{Conclusions and future work} \label{sec.conclusions}
A new AMPC controller, for constrained LTV systems, has been devised to tackle simultaneously both objectives of the dual control problem. The proposed method relies on the partition of the plant's input into a regulatory part and an exciting part. The latter is designed, via a novel MPC-like optimization problem, to persistently excite the system and thereby generate informative enough data for accurate estimates to be obtained. A shortcoming of the proposed approach is that the current estimates might not be able to replace the MPC prediction model if the robust properties are to be maintained (i.e., to account for future changes in the plant), however a variety of methods to verify and ensure that a model update is feasible are presented.

Future work will focus on reducing the conservatism with which the parametric uncertainty is represented (possibly through the implementation of time-varying robust invariant set) and on the extension of the proposed approach to the switching systems framework.

\section*{\refname}
\bibliographystyle{elsarticle-num}

\end{document}